\documentclass[12pt]{iopart}

\usepackage[draft,addedmarkup=colored,deletedmarkup=sout,
authormarkup=superscript,authormarkupposition=left,ulem=normalem]{changes}
\colorlet{Changes@Color}{red}  
\usepackage{todonotes}
\usepackage{graphicx}
\usepackage{epstopdf}
\usepackage{cite}  
\usepackage{amssymb}
\usepackage{bm}
\usepackage{xcolor,xspace}
\usepackage{hyperref}
\hypersetup{colorlinks=True,urlcolor=blue,linkcolor=blue,
citecolor=blue,filecolor=black}
\usepackage{braket}
\usepackage{array}
\usepackage[percent]{overpic}  
\newcolumntype{L}{>{$}l<{$}} 

\newcommand{\ba}{\begin{eqnarray}}
\newcommand{\ea}{\end{eqnarray}}
\newcommand{\bmath}{\begin{mathletters}}
\newcommand{\emath}{\end{mathletters}}
\newcommand{\ban}{\begin{eqnarray*}}
\newcommand{\ean}{\end{eqnarray*}}
\newcommand{\bsub}{\begin{subequations}}
\newcommand{\esub}{\end{subequations}}

\def\ket#1{|#1\rangle}
\def\bra#1{\langle#1|}
\def\beq{\beta_{\rm eq}}
\def\gaeq{\gamma_{\rm eq}}

\begin{document}

\title[Interplay between shape-phase transitions...]{
  Interplay between shape-phase transitions and shape coexistence
  in the Zr isotopes}

\author{N Gavrielov$^1$, A Leviatan$^1$ and F Iachello$^2$}

\address{$^1$ Racah Institute of Physics, The Hebrew University,
Jerusalem 91904, Israel}
\address{$^2$ Center for Theoretical Physics, Sloane Physics Laboratory,
Yale University, New Haven, Connecticut 06520-8120, USA}

\eads{\mailto{noam.gavrielov@mail.huji.ac.il},
\mailto{ami@phys.huji.ac.il} and \mailto{francesco.iachello@yale.edu}}
\vspace{10pt}
\begin{indented}
\item[]September 2019
\end{indented}

\begin{abstract}
  We investigate the evolution of structure in the zirconium isotopes
  where one of the most complex situations encountered in
  nuclear physics occurs. We demonstrate the role of two concurrent 
  types of quantum phase transitions, sharing a common critical point.
  The first type, involves an abrupt crossing of coexisting normal and
  intruder configurations. The second type, involves a gradual
  shape-phase transition within the intruder configuration, changing
  from weakly-deformed to prolate-deformed and finally to gamma-unstable.
  Evidence for this scenario is provided by a detailed comparison with
  experimental data, using a definite algebraic framework.
\end{abstract}

\submitto{\PS}
\ioptwocol

\section{Introduction}\label{sec:intro}

Nuclei in the $Z \approx 40$, $A\approx 100$ region have long 
been recognized to exhibit an abrupt transition from spherical to
deformed ground states and the emergence of shape-coexisting
states~\cite{Cheifetz1970,Federman1977,Federman1979}.
From a shell-model perspective, the sudden onset of deformation at
neutron number 60, has been ascribed to a strong isoscalar proton-neutron
interaction between nucleons occupying the $1g_{9/2}$-$1g_{7/2}$ spin-orbit
partners~\cite{Federman1977,Federman1979,HeydeCas1985,Heyde2011},
producing a crossing of normal and intruder configurations
(the latter arising from the promotion of two protons across the
$Z\!=\!40$ sub-shell gap). These dramatic structural changes
have attracted considerable theoretical and experimental interest.  
In the Zr chain, they have been studied in a variety of
theoretical approaches, including mean-field based methods, both
non-relativistic~\cite{Delaroche2010,Nomura2016} and
relativistic~\cite{Mei2012},
large-scale shell model calculations~\cite{Langanke2009,Petrovici2012}
and the Monte-Carlo shell-model (MCSM)~\cite{Togashi2016}.
The Zr isotopes have been recently the subject of several experimental
investigations~\cite{Chakraborty2013,Browne2015,Kremer2016,Ansari2017,
  Paul2017,Witt2018,Singh2018},
opening the door for understanding the properties
of both yrast and non-yrast states.

Qualitative changes in the ground state properties of
a physical system, induced by a variation of parameters in
the quantum Hamiltonian, are called quantum phase transitions
(QPTs)~\cite{Gilmore1978,Gilmore1979}.
The latter have in recent years become of great interest in a variety
of fields~\cite{Carr}. In nuclei, examples of QPTs are shape changes
within a single configuration, as observed in the neutron number 90
region for Nd-Sm-Gd isotopes~\cite{CejJolCas2010},
and shape coexistence involving multiple configurations,
as observed in nuclei near shell closure, {\it e.g.}, the light
Pb-Hg isotopes~\cite{Heyde2011},
with strong mixing between the configurations.
\begin{figure*}[t]
\centering
\includegraphics[width=0.8\linewidth]{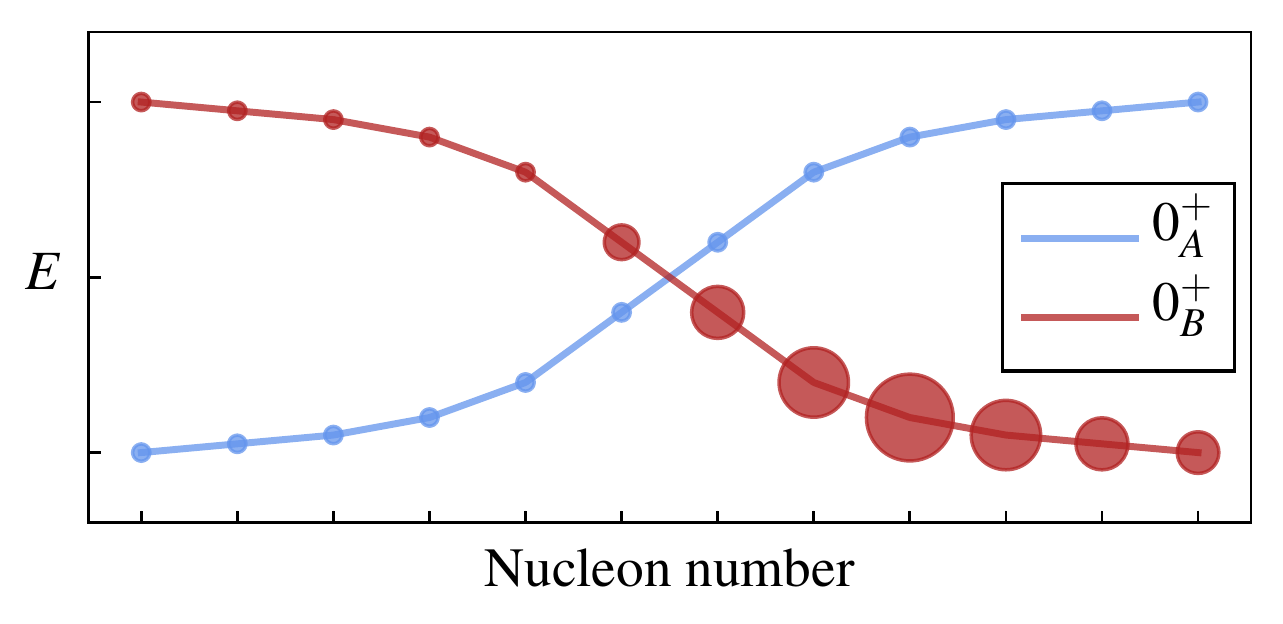}
\caption{Schematic illustration of the scenario of
  intertwined quantum phase transitions.
  The evolution with nucleon number of energies (in arbitrary units) of the
  lowest $0^{+}$ states of two configurations, $A$ and $B$, discloses an
  abrupt crossing. This change in configurations is accompanied by
  gradual changes of shapes (denoted by circles of different size) within
  each configuration.
  \label{fig:iqpt-schem}}
\end{figure*}

In the present work, we show that these
different types of QPTs~\cite{Heyde2004}
play a role in the Zr chain, and that in parallel to an abrupt
swapping of configurations, each configuration maintains its purity
and its own gradual shape-evolution with nucleon number.
This situation, referred to as intertwined
quantum phase transitions~\cite{GavLevIac2019},
gives rise to an intricate interplay between
shape-phase transitions and shape coexistence in nuclei.

The notion of intertwined quantum phase transitions is illustrated
schematically in Fig.~\ref{fig:iqpt-schem}, where a sudden crossing of
two configurations vs. nucleon number is superimposed on progressive
shape-changes in each configuration. In what follows, we provide
evidence for such a scenario in the Zr isotopes by means of a detailed
comparison with experimental data, analyzed in
a physically transparent symmetry-based framework,
the interacting boson model (IBM)~\cite{IBM}.
After a brief review in Section~2 of the model and its extensions
to accommodate configuration mixing,
we apply the formalism to the Zr chain in Section~3 and present
both a quantum and a classical analysis. 
In Section~4 we discuss the evolution of the quantum spectra with nucleon
number, identifying the underlying multiple QPTs.
In Section~5 we present the corresponding evolution of order parameters
and related observables, including $E2$ transition rates, isotope shifts
and two-neutron separation energies. Concluding remarks are collected
in Section~6.

\section{QPTs in the IBM and its extensions}

The IBM describes low lying quadrupole states in even even nuclei in terms of
a system of monopole ($s$) and quadrupole ($d$) bosons, representing valence
nucleon pairs~\cite{IBM,IacTal1987}.
For a single shell-model configuration space,
the total number of bosons is conserved and is fixed by the
microscopic interpretation to be $N\!=\!N_{\pi}+N_{\nu}$,
where $N_{\pi}$ ($N_{\nu}$) is the number of proton (neutron)
particle or hole pairs counted from the nearest closed shell.
In its simplest version, the IBM has U(6) as a spectrum generating
algebra and exhibits three dynamical symmetry (DS) limits with
leading subalgebras: U(5), SU(3) and SO(6), whose analytic solutions
resemble known paradigms of collective motion: spherical vibrator,
axially-symmetric and $\gamma$-soft deformed rotors, respectively.
A geometric visualization of the IBM is obtained by an energy surface,
\ba
E_{N}(\beta,\gamma) &=&
\bra{\beta,\gamma; N} \hat{H} \ket{\beta,\gamma ; N} ~,
\label{enesurf}
\ea 
defined by the expectation value of the Hamiltonian in
the following coherent (intrinsic) state~\cite{GinKir1980,Diep1980},
\numparts 
\label{eq:coherent}
\ba
&&  \ket{\beta,\gamma;N} = (N!)^{-1/2}(b^\dagger_c)^N\ket{0}~,\\
&&b^\dagger_c = (1+\beta^2)^{-1/2}[\beta \cos \gamma~d^\dagger_0 \nonumber
\nonumber  \\
&&  \qquad\qquad
  + \beta \sin \gamma (d^\dagger_2 + d^\dagger_{-2})/\sqrt{2} + s^\dagger]~.
\ea
\endnumparts
Here $(\beta,\gamma)$ are
quadrupole shape parameters whose values, $(\beta_{\rm eq},\gamma_{\rm eq})$, 
at the global minimum of $E_{N}(\beta,\gamma)$ define the equilibrium 
shape for a given Hamiltonian.
For two body interactions,
the shape can be spherical $(\beta_{\rm eq} \!=\!0)$ or 
deformed $(\beta_{\rm eq} \!>\!0)$ with $\gamma_{\rm eq} \!=\!0$ (prolate), 
$\gamma_{\rm eq} \!=\!\pi/3$ (oblate), or $\gamma$-independent.

The dynamical symmetries correspond to possible phases of the system.
QPTs can be studied in the IBM using an Hamiltonian $\hat{H}(\xi)$
which interpolates between the DS limits (phases) by varying its control
parameters $\xi$. The related energy surface
$E_{N}(\beta,\gamma;\xi)$ serves as the Landau potential,
whose topology determines
the type of phase transition (Ehrenfest classification).
The order parameter is taken to be the expectation value
of the $d$-boson number operator, $\hat{n}_d$, in the ground state
\ba
\frac{\braket{\hat{n}_d}}{N}\approx
\frac{\beta_{\rm eq}^2}{1+\beta_{\rm eq}^2} ~,
\label{order-p}
\ea
which in turn is related to
the expectation value in $\ket{\beq,\gaeq;N}$, hence to
the equilibrium deformation, $\beq$.
The dependence of $\beq$ on $\xi$, discloses the order
of the transition.

QPTs involving a single configuration have been studied extensively
in the IBM framework~\cite{CejJolCas2010,Diep1980,CejJol2009,Iac2011}.
A typical Hamiltonian frequently used in such studies,
has the form~\cite{Warner1983,Lipas1985},
\ba
  \label{eq:H-single-2}
  \hat H(\epsilon_d,\kappa,\chi) =
  \epsilon_d\, \hat n_d + \kappa\, \hat Q_\chi \cdot \hat Q_\chi ~,
\ea
where the quadrupole operator is given by
\ba
\hat Q_\chi =
d^\dag s+s^\dag \tilde d \!+\!\chi (d^\dag \times \tilde d)^{(2)} ~.
\label{Quad}
\ea
Here $\tilde{d}_{m} = (-1)^{m}d_{-m}$ and standard notation of angular
momentum coupling is used. The associated Landau potential reads
\begin{eqnarray} \label{eq:ener-func-single}
  &&E_{N}(\beta,\gamma;\epsilon_d,\kappa,\chi) =
  \nonumber\\
  &&\quad 5\kappa\, N + \frac{N\beta^2}{1+\beta^2} \left [
    \,\epsilon_d + \kappa (\chi^2-4)\,\right ]
  \nonumber\\
  &&\quad + \frac{N(N-1)\beta^2}{(1+\beta^2)^2}\kappa
  \left [\, 4 - 4\bar{\chi}\beta\,\Gamma
    + \bar{\chi}^2\beta^2\,\right ] ,\qquad
 \end{eqnarray} 
where 
$\bar{\chi} = \sqrt{\frac{2}{7}}\chi$ and $\Gamma = \cos 3\gamma$.
The control parameters $(\epsilon_d,\kappa,\chi)$ in
Eq.~(\ref{eq:H-single-2}),
interpolate between the U(5), SU(3) and SO(6) DS limits, which are
reached for $(\kappa\!=\!0)$,
$(\epsilon_d\!=\!0,\chi\!=\!-\frac{\sqrt{7}}{2})$
and $(\epsilon_d\!=\!0,\chi\!=\!0)$, respectively. The U(5)-SU(3)
transition is found to be first-order, the U(5)-SO(6) transition is
second order and the SU(3)-SO(6) transition is a crossover. 
\begin{figure*}[t]
\centering
\includegraphics[width=0.85\linewidth]{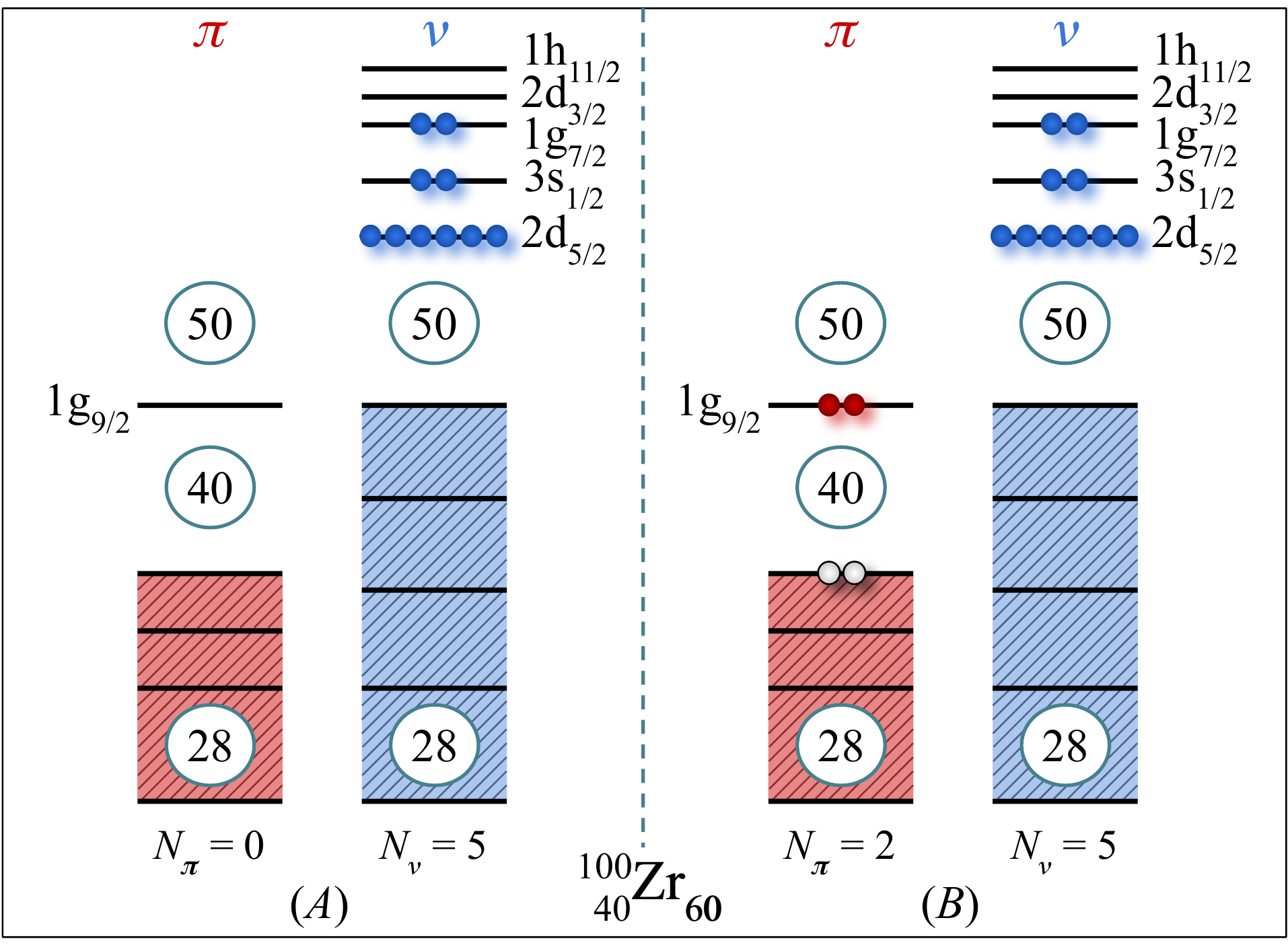}
\caption{Schematic representation of the two coexisting shell-model
configurations ($A$ and $B$) for $^{100}_{40}$Zr$_{60}$.
The corresponding numbers of proton bosons ($N_{\pi}$) and neutron bosons
($N_{\nu}$), relevant to the IBM-CM, are listed for each configuration.
\label{fig:shells}}
\end{figure*}

An extension of the IBM to include intruder excitations
is based on associating the different shell-model
spaces of 0p-0h, 2p-2h, 4p-4h,$\dots$ particle-hole excitations,
with the corresponding boson spaces comprising of
$N,\, N\!+\!2,\, N\!+\!4,\ldots$ bosons, which are
subsequently mixed. The resulting interacting boson model with
configuration mixing (IBM-CM)~\cite{Duval1981,Duval1982}
has been used extensively for describing
configuration-mixed QPTs and coexistence phenomena in
nuclei~\cite{Duval1981,Duval1982,Sambataro1982,Duval1983,Bijker2006,
Fossion2003,Frank2006,Ramos2011,Ramos2014,Ramos2015,Nomurajpg2016,
LevGav2018}.
In this case, the quantum Hamiltonian has a matrix form~\cite{Frank2006}
\begin{eqnarray} \label{eq:H-mult-mat}
\hat{H}(\xi_A,\xi_B,\omega) =
\left [
\begin{array}{cc}
\hat{H}_{A}(\xi_A) & \hat{W}(\omega) \\ 
\hat{W}(\omega) & \hat{H}_{B}(\xi_B)
\end{array}
\right ] ~,
\end{eqnarray}
where the index $A$, $B$ denote the two configurations.
The Hamiltonian $\hat{H}_{A}(\xi_A)$
acts on the $A$ (normal) configuration,
corresponding to the valence space and
$\hat{H}_{B}(\xi_B)$ on the $B$ (intruder) configuration,
corresponding to
the core-excited excitations. The 
$\hat{W}(\omega)$ term mixes both spaces.
When two configurations coexist,
the energy surface becomes a matrix,
\begin{eqnarray}\label{eq:potential-mat}
E(\beta,\gamma) =
\left [
\begin{array}{cc}
E_A(\beta,\gamma;\xi_A) & \Omega(\beta,\gamma;\omega) \\ 
\Omega(\beta,\gamma;\omega) & E_B(\beta,\gamma;\xi_B)
\end{array}
\right ] ,\quad
\end{eqnarray} 
whose entries are the matrix elements of the
corresponding terms
in the Hamiltonian~(\ref{eq:H-mult-mat}), between 
the intrinsic states~(\ref{eq:coherent}) of the two
configurations, with appropriate boson numbers.
Diagonalization of this two-by-two
matrix produces the so-called eigen-potentials,
$E_{\pm}(\beta,\gamma)$~\cite{Frank2006,Frank2004,Hellemans2007}.

As the control parameters $(\xi_A,\xi_B,\omega)$ in the
Hamiltonian~(\ref{eq:H-mult-mat}) are varied,
the two coexisting configurations can exchange roles, and their
individual shapes can evolve. Usually the latter quantum
shape-phase transitions are masked by the strong mixing between the
two configurations. In what follows,
we show that the Zr isotopes are exceptional in the sense that
the crossing is abrupt, the separate configurations retain their
purity before and after the crossing, and the shape evolution of the
intruder configuration can be cast in terms of its own phase transition.

\section{The IBM-CM in the Zr chain}

To describe the $_{40}$Zr isotopes in the IBM-CM framework, requires
a choice of Hilbert space, Hamiltonian and transition operators.
Similar to a calculation done for the $_{42}$Mo isotopes
in~\cite{Sambataro1982}, we consider
$_{40}^{90}$Zr$_{50}$ as a core and valence neutrons in the 50-82 major
shell. The normal $A$-configuration corresponds to having
no active protons above $Z\!=\!40$ sub-shell gap,
and the intruder $B$-configuration corresponds to two-proton excitation
from below to above this gap, creating 2p-2h states.
According to the usual boson-counting,
the corresponding bosonic configurations have proton bosons
$N_{\pi}\!=\!0$ for the normal configuration and $N_{\pi}\!=\!2$
for the intruder configuration. Both configurations have neutron bosons
$N_{\nu}\!=\!1,~2,\ldots,~8$ for neutron number 52-66,
and $\bar{N}_{\nu}\!=\!7,\,6$ for 68-70, where
the bar over the number indicates that these are hole bosons.
Altogether, the IBM-CM model space, employed in the current study,
consists of a $[N]\oplus[N+2]$ boson space with total boson number
$N\!=\!1,2,\ldots,8$ for $^{92-106}$Zr and $\bar{N}\!=\!7,6$ for
$^{108,110}$Zr. These two configurations are shown in
Fig.~\ref{fig:shells}, for $^{100}_{40}$Zr$_{60}$.

We write the Hamiltonian not in the matrix form of
Eq.~(\ref{eq:H-mult-mat}), but rather in the equivalent form
\begin{equation} \label{eq:H-mult}
\hat{H}=\hat{H}_{A}^{(N)}+\hat{H}_{B}^{(N+2)}+\hat{W}^{(N,N+2)} ~.
\end{equation}
Here $\hat{\cal O}^{(N)}=\hat{P}_{N}^{\dag }\hat{\cal O}\hat{P}_{N}$ and
$\hat{\cal O}^{(N,N^{\prime })}=
\hat{P}_{N}^{\dag }\hat{\cal O}\hat{P}_{N^{\prime }}$, 
for an operator $\hat{\cal O}$, with $\hat{P}_{N}$, a projection operator 
onto the $[N] $ boson space. The Hamiltonian $\hat{H}_{A}^{(N)}$ represents
the normal ($N$ boson space) configuration and $\hat{H}_{B}^{(N+2)}$
represents the intruder configuration ($N\!+\!2$ boson space).
The explicit form of these Hamiltonians is given by
\numparts
\label{H-AB}
\ba
\hat H_A = & \hat H(\epsilon^{(A)}_d,\kappa^{(A)},\chi)~,
\label{eq:HA}
\\
\hat{H}_{B} = & \hat H(\epsilon^{(B)}_d,\kappa^{(B)},\chi)
+ \kappa^{\prime(B)} \hat L \cdot \hat L 
+ \Delta_p\; .\;\;
\label{eq:HB}
\ea
\endnumparts
They involve terms similar to those of the single-configuration
Hamiltonian of Eq.~(\ref{eq:H-single-2}). $\hat{H}_B$ of
Eq.~(\ref{eq:HB}), contains an additional rotational term and
$\Delta_p$ is an off-set between the normal and intruder configurations,
where the index $p$ denotes the fact that this is a proton excitation.
The mixing term in Eq.~(\ref{eq:H-mult}) has the
form~\cite{IBM,Duval1981, Duval1982}
\begin{equation}
\label{eq:mixing}
\hat{W} = \omega\,[\,( d^{\dag }\times d^{\dag })^{(0)}
+ \,(s^{\dag })^2\,] + {\rm H.c.} ~,\quad
\end{equation}
where H.c. stands for Hermitian conjugate.
The resulting eigenstates $\ket{\Psi;L}$ 
with angular momentum $L$, are linear combinations of the
wave functions, $\Psi_A$ and $\Psi_B$,
in the two spaces $[N]$ and $[N+2]$,
\ba
\label{eq:wf}
\ket{\Psi; L} = a\ket{\Psi_A; [N], L} + b\ket{\Psi_B; [N\!+\!2], L}\, ,\;
\ea
with $a^{2} + b^{2} = 1$. The above decomposition reflects
the normal-intruder mixing in the state considered.
\begin{figure*}[t]
\centering
\includegraphics[width=0.85\linewidth]{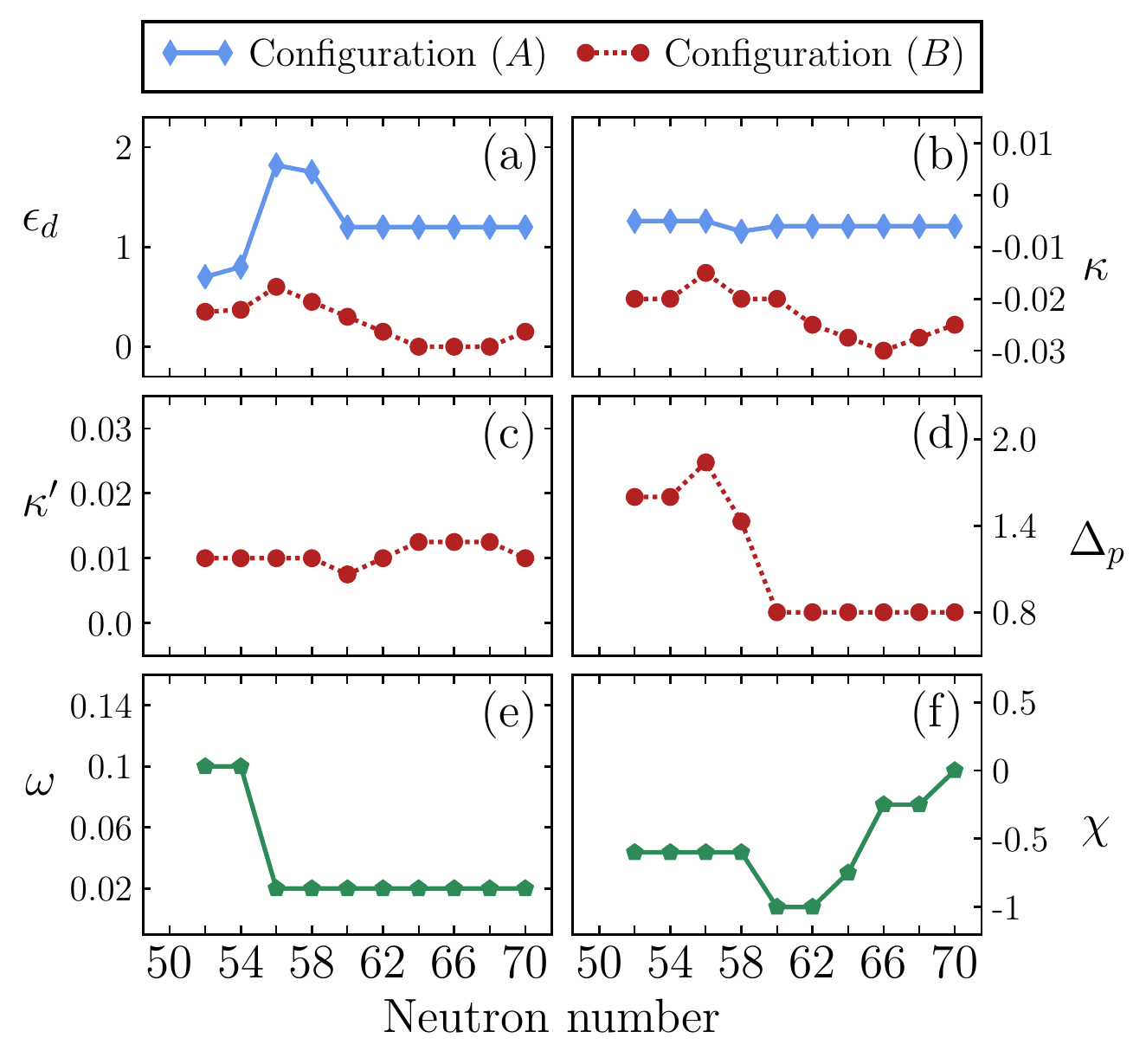}
  \caption{Parameters of the IBM-CM Hamiltonians, Eqs.~\eref{eq:HA},
    \eref{eq:HB}, \eref{eq:mixing}, are in MeV and the parameter $\chi$
    of Eq.~(\ref{Quad}), is dimensionless.
\label{fig:params}}
\end{figure*} 

Adapted to two configurations, the $E2$ operator reads 
\begin{equation}
\hat{T}(E2) = e^{(A)}\hat Q^{(N)}_{\chi} + e^{(B)}\hat Q^{(N+2)}_{\chi},
\label{TE2}
\end{equation} 
with~$\hat{Q}_{\chi}^{(N)}\!=\!\hat{P}_{N}^{\dag}\hat{Q}_{\chi}\hat{P}_{N}$,
$\hat{Q}_{\chi}^{(N+2)}\!=\!P_{N+2}^{\dag }\hat{Q}_{\chi}\hat{P}_{N+2}$,
and $\hat{Q}_{\chi}$, defined in Eq.~(\ref{Quad}), 
is the same quadrupole operator appearing in the Hamiltonian.
In Eq.~(\ref{TE2}), $e^{(A)}$ and $e^{(B)}$ are the boson effective charges
for the configurations $A$ and $B$, respectively.

A geometric interpretation is obtained by means of the matrix 
$E(\beta,\gamma)$, Eq.~(\ref{eq:potential-mat}), with 
entries $E_{A}(\beta,\gamma)\!=\!
\braket{\beta,\gamma;N|\hat{H}_A|\beta,\gamma;N}$,
$E_{B}(\beta,\gamma) \!=\!
\braket{\beta,\gamma;N\!+\!2|\hat{H}_B|\beta,\gamma;N\!+\!2}$ and
$\Omega(\beta,\gamma) \!=\!
\braket{\beta,\gamma;N|\hat{W}|\beta,\gamma;N\!+\!2}$.
These entries involve the expectation values
of the Hamiltonians
$\hat{H}_A$~(\ref{eq:HA}) and $\hat{H}_B$~(\ref{eq:HB}),
in the intrinsic states~(\ref{eq:coherent}), with $N$ and $N \!+\! 2$
bosons respectively, and a non-diagonal matrix element of
the mixing term $\hat{W}$~(\ref{eq:mixing}), between them.
The explicit expressions are found to be
\numparts
\begin{eqnarray}
E_A(\beta,\gamma) & = 
E_{N}(\beta,\gamma;\epsilon^{(A)}_d,\kappa^{(A)},\chi)~,
\label{EA}
\\ [2mm]
E_B(\beta,\gamma)  & = 
E_{N+2}(\beta,\gamma;\epsilon^{(B)}_d,\kappa^{(B)},\chi)
\nonumber\\
& \quad + 6\kappa^{\prime(B)}\frac{(N+2)\beta^2}{1+\beta^2}
+ \Delta_p ~,
\label{EB}
\\[2mm]
\Omega(\beta,\gamma) & = 
\textstyle{\frac{\sqrt{(N+2)(N+1)}}{1+\beta^2}}
\omega\Bigg(1 + \frac{1}{\sqrt{5}}\beta^2\Bigg)~,
\label{EW}
\end{eqnarray}
\endnumparts
where the surfaces on the right-hand-side of Eqs.~(\ref{EA})-(\ref{EB})
are obtained from Eq.~(\ref{eq:ener-func-single}).
\begin{figure*}[t!]
\centering
\begin{overpic}[width=0.33\linewidth,height=0.24\textheight,
keepaspectratio]{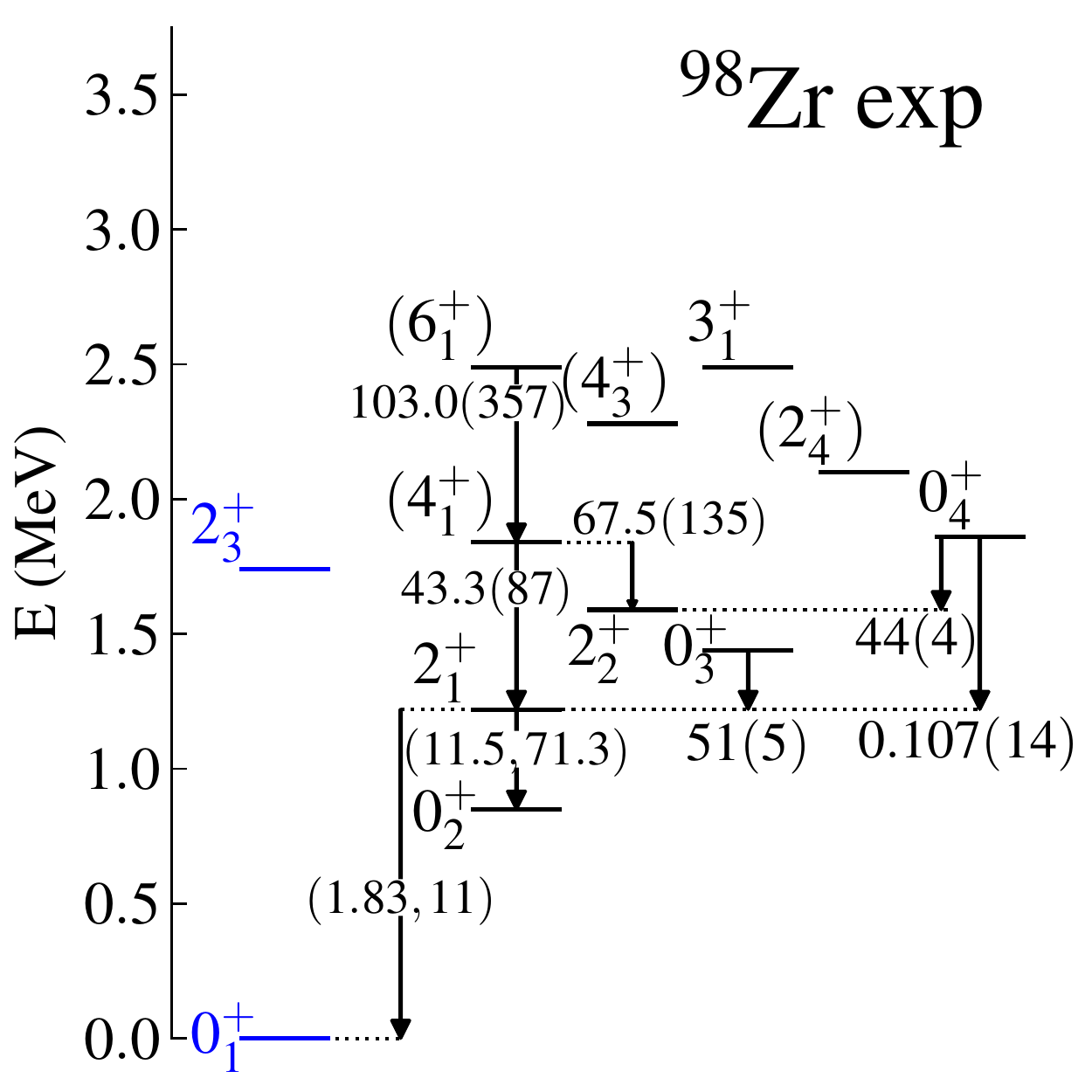}
\put (80,75) {\large(a)}
\end{overpic}
\begin{overpic}[width=0.33\linewidth,height=0.24\textheight,
keepaspectratio]{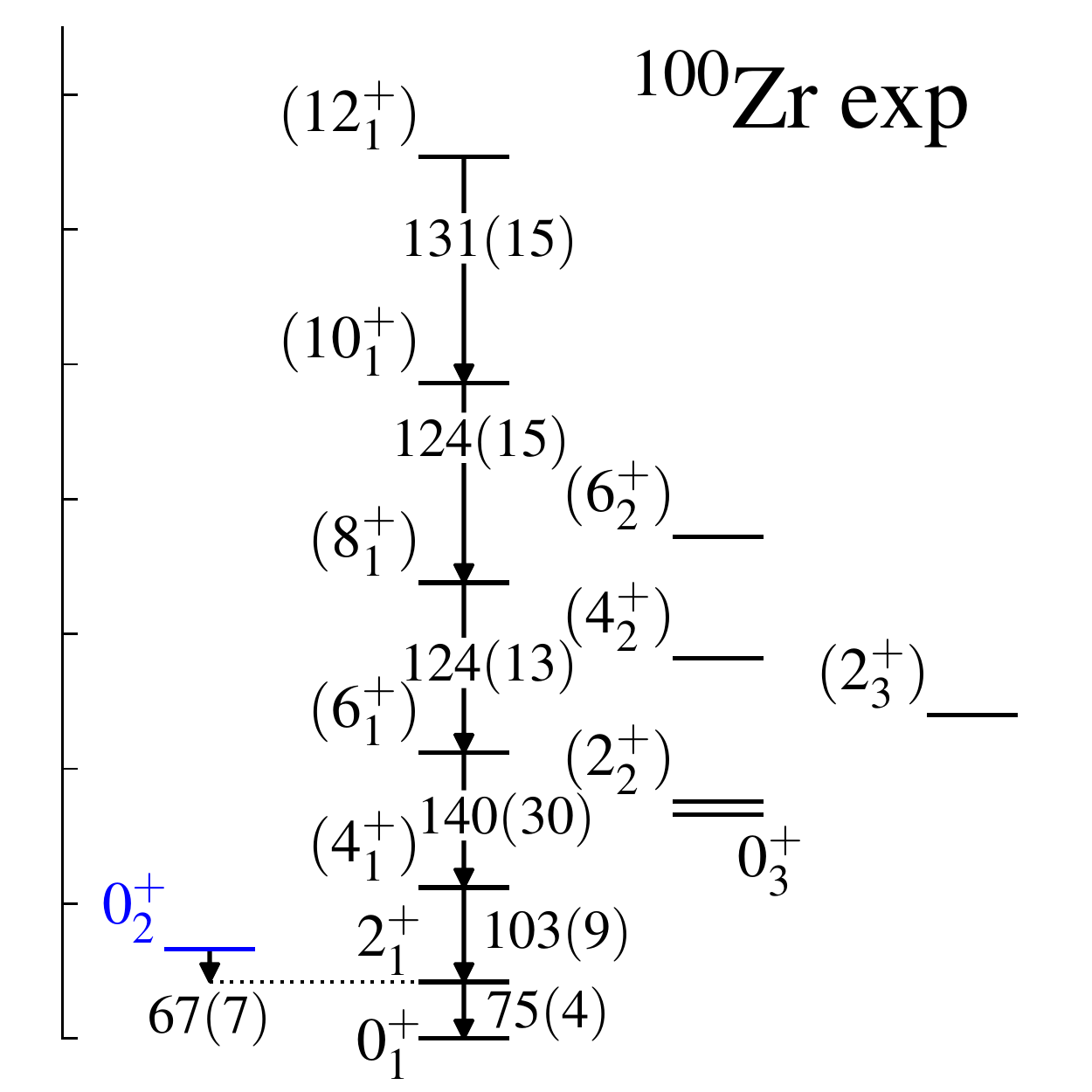}
\put (80,75) {\large(c)}
\end{overpic}
\begin{overpic}[width=0.33\linewidth,height=0.24\textheight,
keepaspectratio]{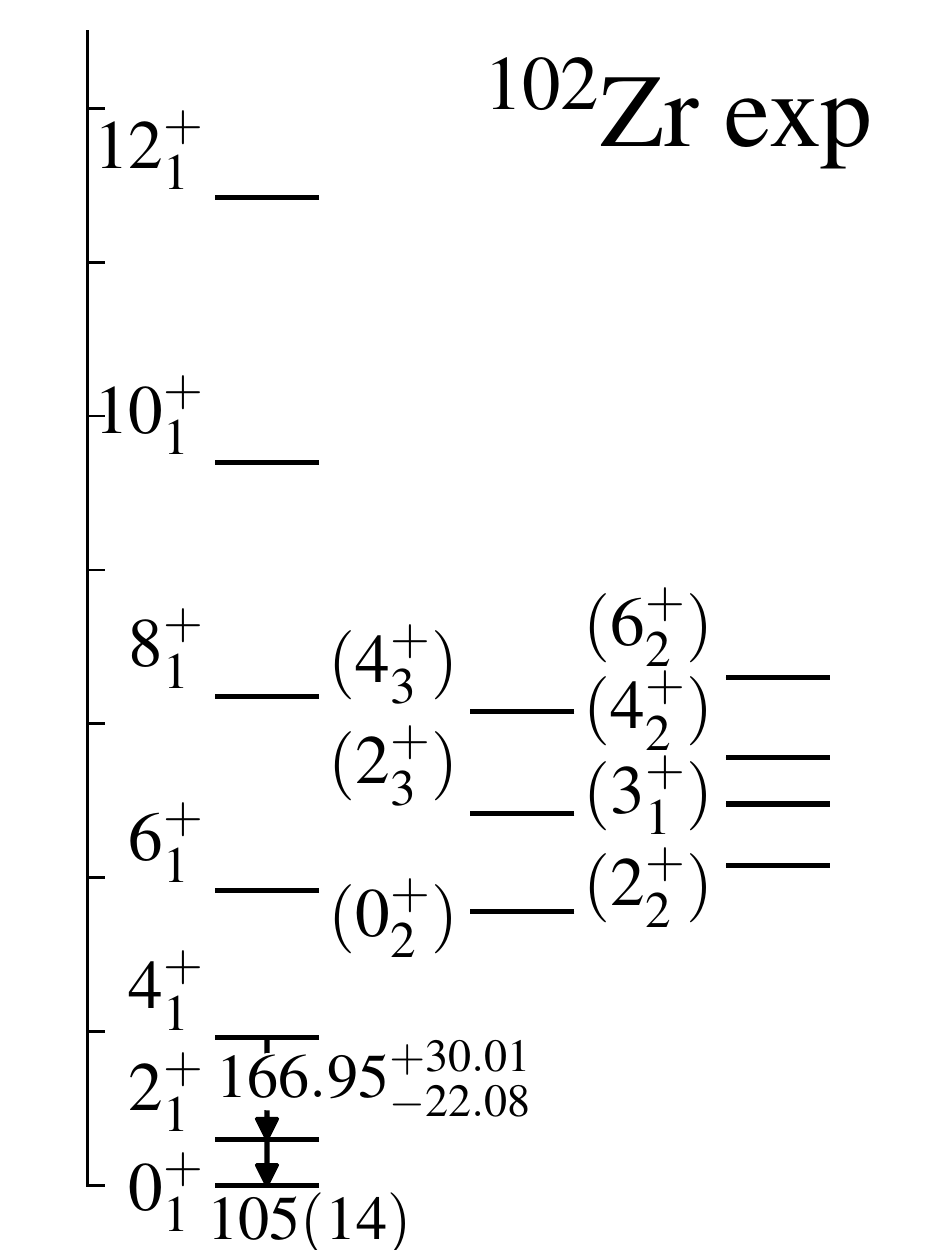}
\put (60,75) {\large(e)}
\end{overpic}\\
\begin{overpic}[width=0.33\linewidth,height=0.24\textheight,
keepaspectratio]{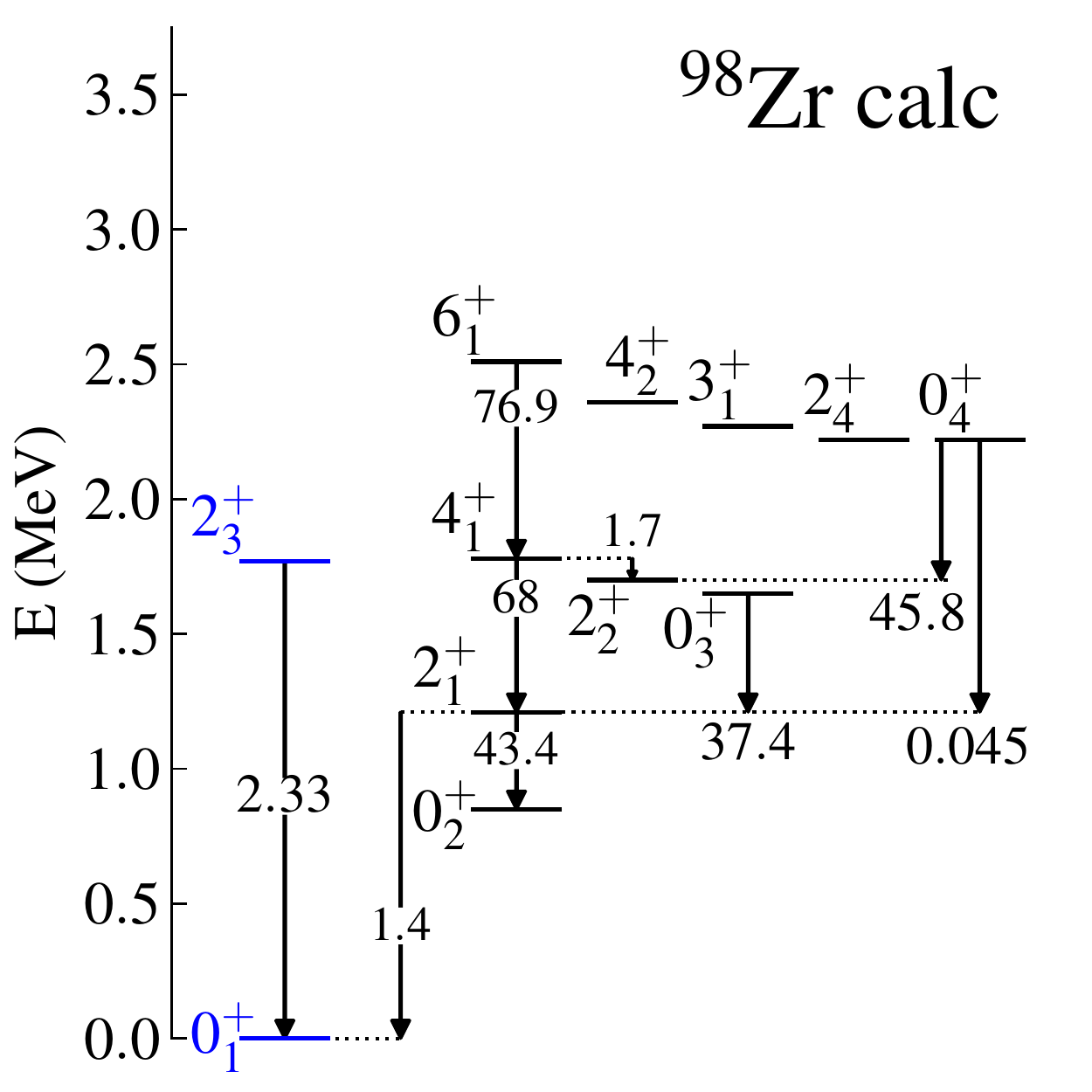}
\put (80,75) {\large(b)}
\end{overpic}
\begin{overpic}[width=0.33\linewidth,height=0.24\textheight,
keepaspectratio]{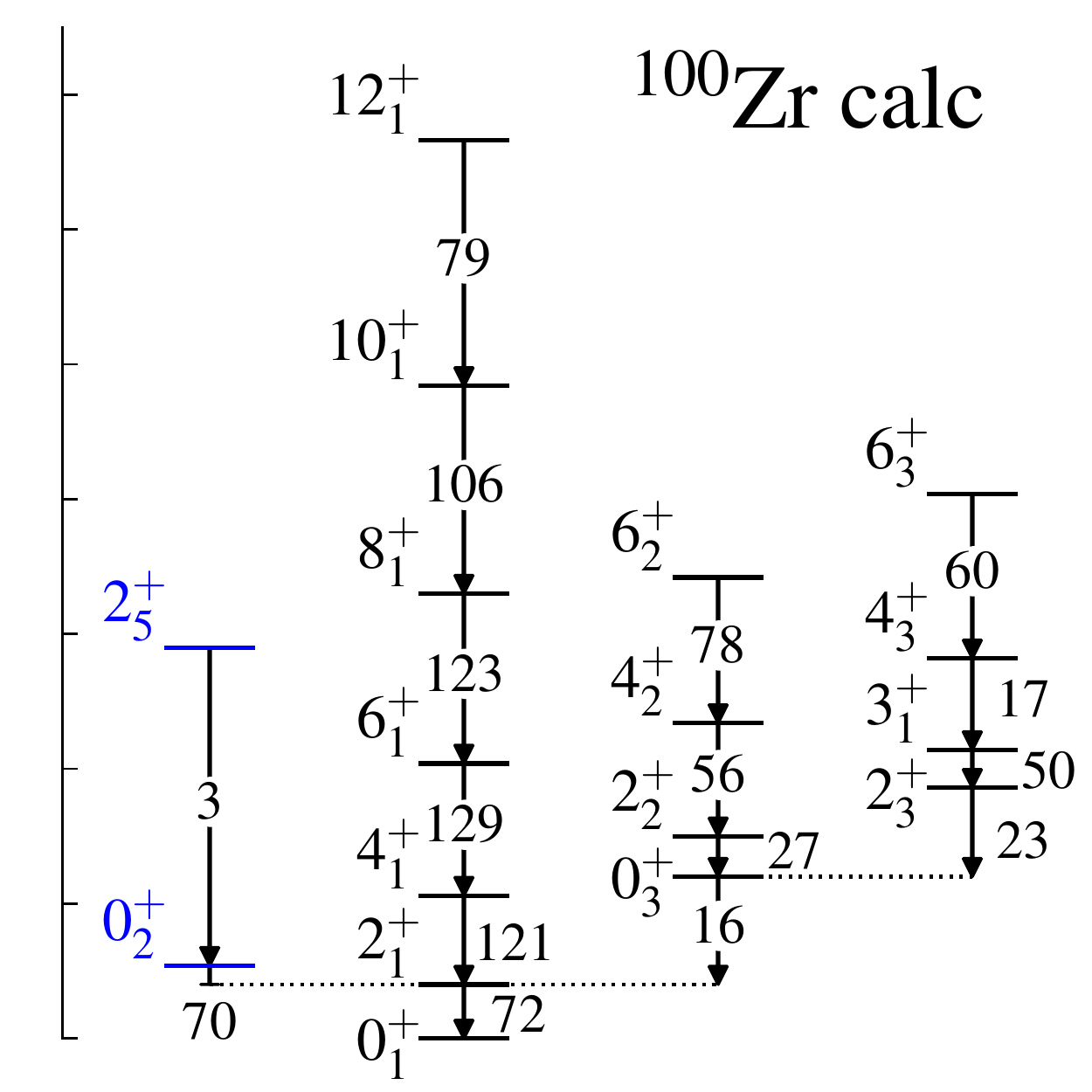}
\put (80,75) {\large(d)}
\end{overpic}
\begin{overpic}[width=0.33\linewidth,height=0.24\textheight,
keepaspectratio]{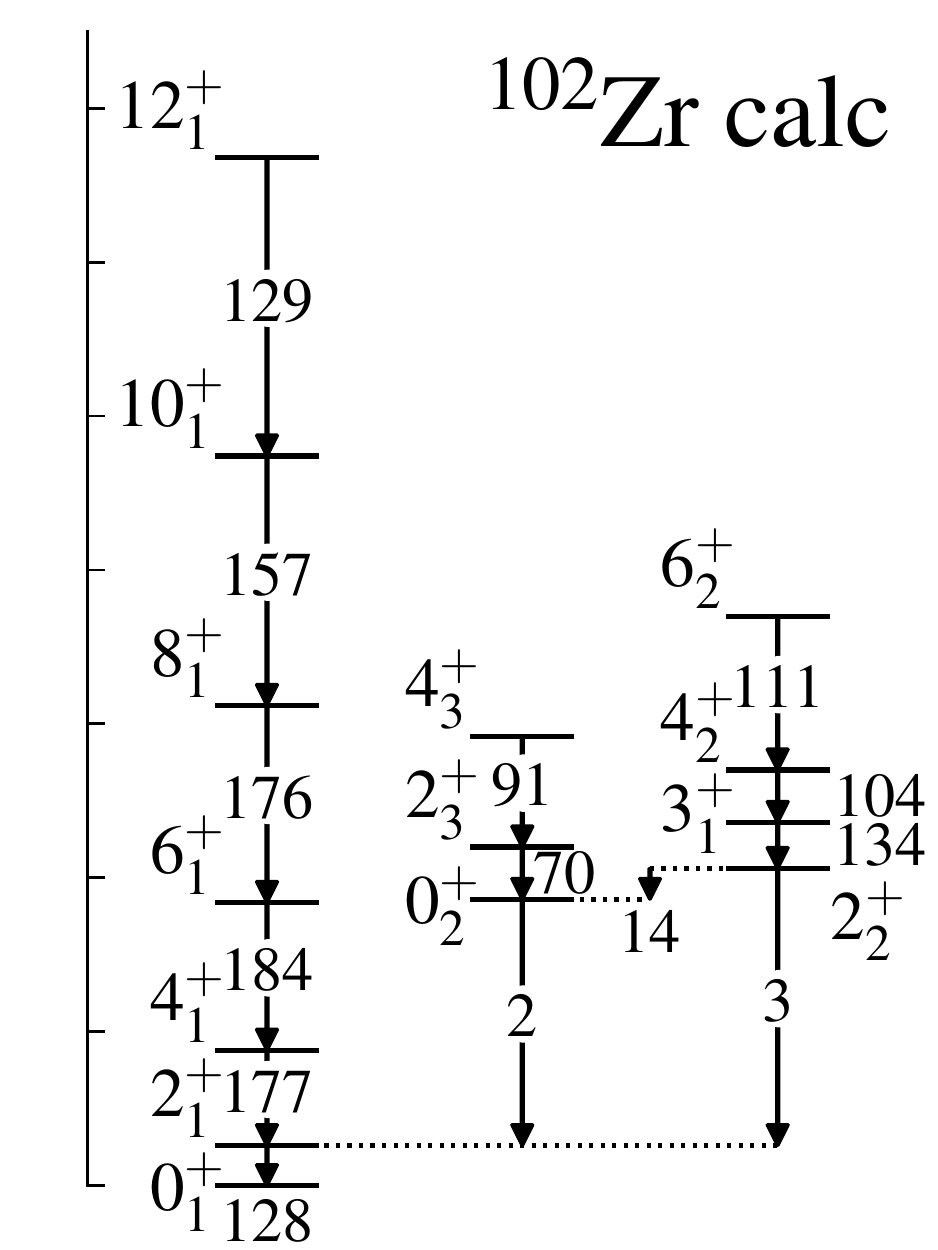}
\put (60,75) {\large(f)}
\end{overpic}
\caption{\label{fig:spectrum} \small
  Experimental~\cite{Ansari2017,Witt2018,Singh2018,ensdf} (top row) and
  calculated (bottom row) energy levels in MeV and $E2$ rates in W.u.
  for $^{98}$Zr [panels (a)-(b)], $^{100}$Zr [panels (c)-(d)] and
  $^{102}$Zr [panels (e)-(f)]. The levels ($0^{+}_1,\,2^{+}_3$)
  in $^{98}$Zr and ($0^{+}_2,\,2^{+}_5$) in$^{100}$Zr are dominated by
  the normal ($A$) configuration. All other levels shown are dominated by
  the intruder ($B$) configuration. Assignments are based on the
  decomposition of Eq.~(\ref{eq:wf}).}
\end{figure*}

\section{Quantum and classical analyses}

A first step in a quantum analysis, involves a numerical
diagonalization of the IBM-CM Hamiltonian,
and evaluating matrix elements of the
$E2$ operator~(\ref{TE2}) between its eigenstates.
The parameters of these operators are determined from a combined fit to
the data on spectra and $E2$ transitions. The calculated observables
are then compared with the measured values.

The adapted fitting procedure is similar to that used
in~\cite{Sambataro1982,Duval1983,Bijker2006,
Fossion2003,Frank2006,Ramos2011,Ramos2014,Ramos2015}.
We allow a gradual change of the parameters between adjacent isotopes,
but take into account the proposed shell-model interpretation for the
structure evolution in this
region~\cite{Federman1977,Federman1979,HeydeCas1985}.
The Hamiltonian parameters used are displayed in Fig.~\ref{fig:params}
and are consistent with those of previous calculations in this mass
region~\cite{Sambataro1982,Duval1983,Bijker2006}.
A~symmetry about mid-shell, at neutron number 66,
was imposed on all parameters (except~$\chi$), in accord with
 microscopic aspects of the IBM~\cite{IacTal1987}.
Apart from some fluctuations due to 
the subshell closure at neutron number 56
(the filling by the neutrons of the $2d_{5/2}$ orbital),
the values of the parameters are 
a smooth function of neutron number and, in some cases, a constant.
A notable exception is the sharp decrease by 1~MeV of
the energy off-set parameter $\Delta_p$
beyond neutron number 56. Such a behavior was observed for the
Mo and Ge chains~\cite{Sambataro1982,Duval1983,Bijker2006} and, 
as noted in~\cite{Sambataro1982}, it reflects the effects of the
isoscalar residual interaction, $V_{pn}$, between protons and neutrons
occupying the partner orbitals $1g_{9/2}$ and $1g_{7/2}$,
which is the established mechanism for descending cross shell-gap
excitations and onset of deformation in this
region~\cite{Federman1979,HeydeCas1985}.
The trend in $\Delta_p$ agrees with shell model estimates for
the monopole correction of $V_{pn}$~\cite{Heyde1987}.
The mixing parameter $\omega$~(\ref{eq:mixing}) is determined from
$E2$ transitions between configurations, and is
kept constant except for neutron numbers 52-54,
where the normal configuration space is small ($N\!=\!1,2$).
In general, the underlying physics in the current IBM study is similar
to that of Refs.~\cite{Federman1979,HeydeCas1985}, which albeit use
a different formal language, in which the lowering in energy and
developed collectivity of the intruder configuration are governed
by the relative magnitude
of $V_{pn}$ (especially its monopole and quadrupole components)
and the energy gaps between spherical
shell-model states near shell and subshell closures.
A more direct relation between the two approaches necessitates
a proton-neutron version of the IBM.
The boson effective charges in the $E2$ operator~(\ref{TE2}), 
$e^{(A)}=0.9$ and $e^{(B)}=2.24$ $({\rm W.u.})^{1/2}$,
are determined from the ${2^+\!\to\!0^+}$
transitions within each configuration
and $\chi$ is the same parameter as in the Hamiltonian,
shown in Fig.~\ref{fig:params}. Fine-tuning the parameters for
individual isotopes can improve the fit, but the main conclusions of the
analysis, to be reported below, are not changed.

The calculations describe the experimental data in the
entire range $^{92-110}$Zr very well.
A~full account is given in~\cite{gavrielov-thes}. Here we show
only three examples, $^{98}$Zr, $^{100}$Zr and $^{102}$Zr, where
a~first-order shape-phase transition takes place,
accompanied by a crossing of the normal and intruder configurations.
$^{98}$Zr, in Figs.~\ref{fig:spectrum}(a)
and \ref{fig:spectrum}(b),
has a spherical [U(5)-like] ground state configuration ($A$)
and a weakly-deformed [U(5)-perturbed] excited configuration ($B$).
$^{100}$Zr is near the critical point of both
types of phase transitions,
and yet our description of energy levels and $B(E2)$ 
values is excellent; see
Fig.~\ref{fig:spectrum}(c) and~\ref{fig:spectrum}(d). 
The ground state band, has now become configuration~($B$), 
and appears to have features of the so-called X(5) 
symmetry~\cite{Iac2001}, while 
the spherical configuration~($A$) has now become the excited band
$0^+_2 $. $^{102}$Zr, in Fig.~\ref{fig:spectrum}(e)
and ~\ref{fig:spectrum}(f), exhibits well developed deformed
[SU(3)-like] rotational bands assigned to configuration~($B$).
States of configuration~($A$) have shifted to higher energies.
\begin{figure*}[t]
\centering
\includegraphics[width=0.95\linewidth]{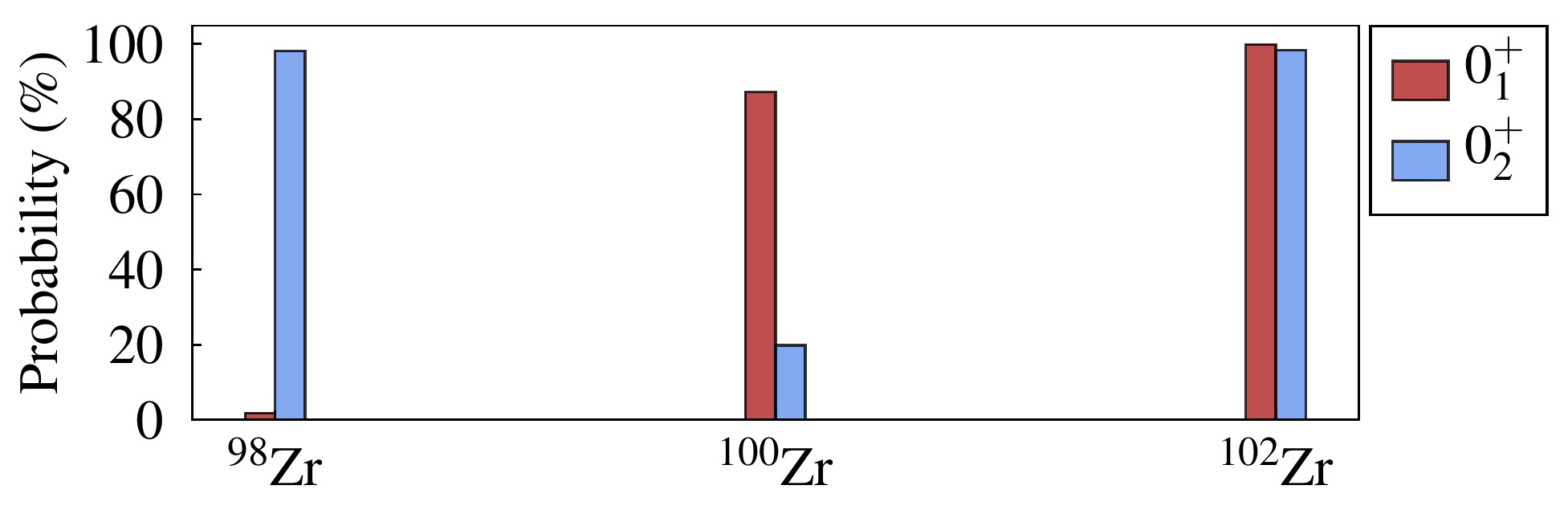}
\caption{Percentage of the wave functions within the intruder
  $B$-configuration [the $b^2$ probability in Eq.~(\ref{eq:wf})],
  for the ground ($0^{+}_1$) and excited ($0^{+}_2$) states in
  $^{98}$Zr, $^{100}$Zr and $^{102}$Zr.
\label{fig:decomp}}
\end{figure*} 
\begin{figure*}[t]
\centering
\begin{overpic}[width=0.19\linewidth]{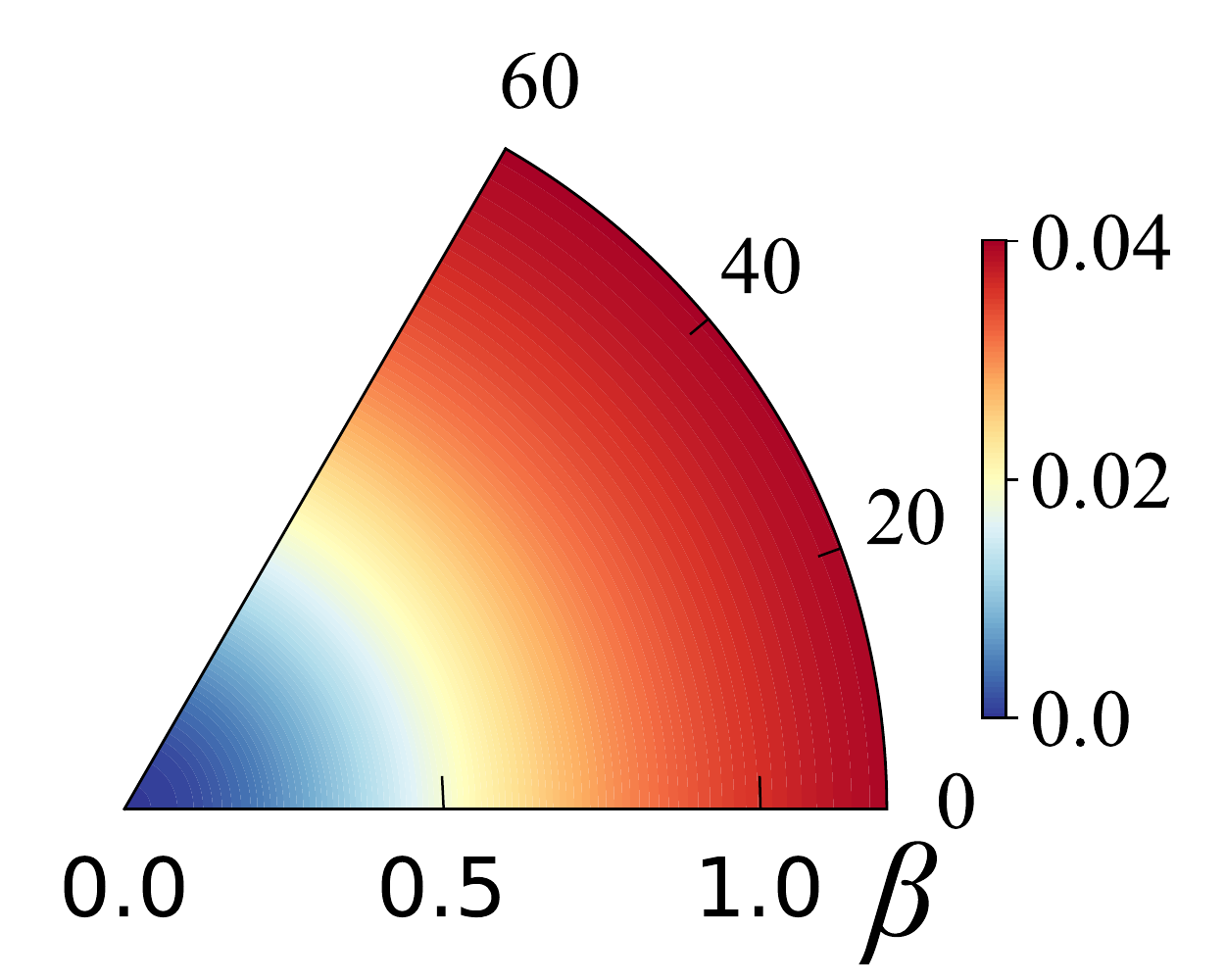}
\put (0,60) {\large $^{92}$Zr}
\end{overpic}
\begin{overpic}[width=0.19\linewidth]{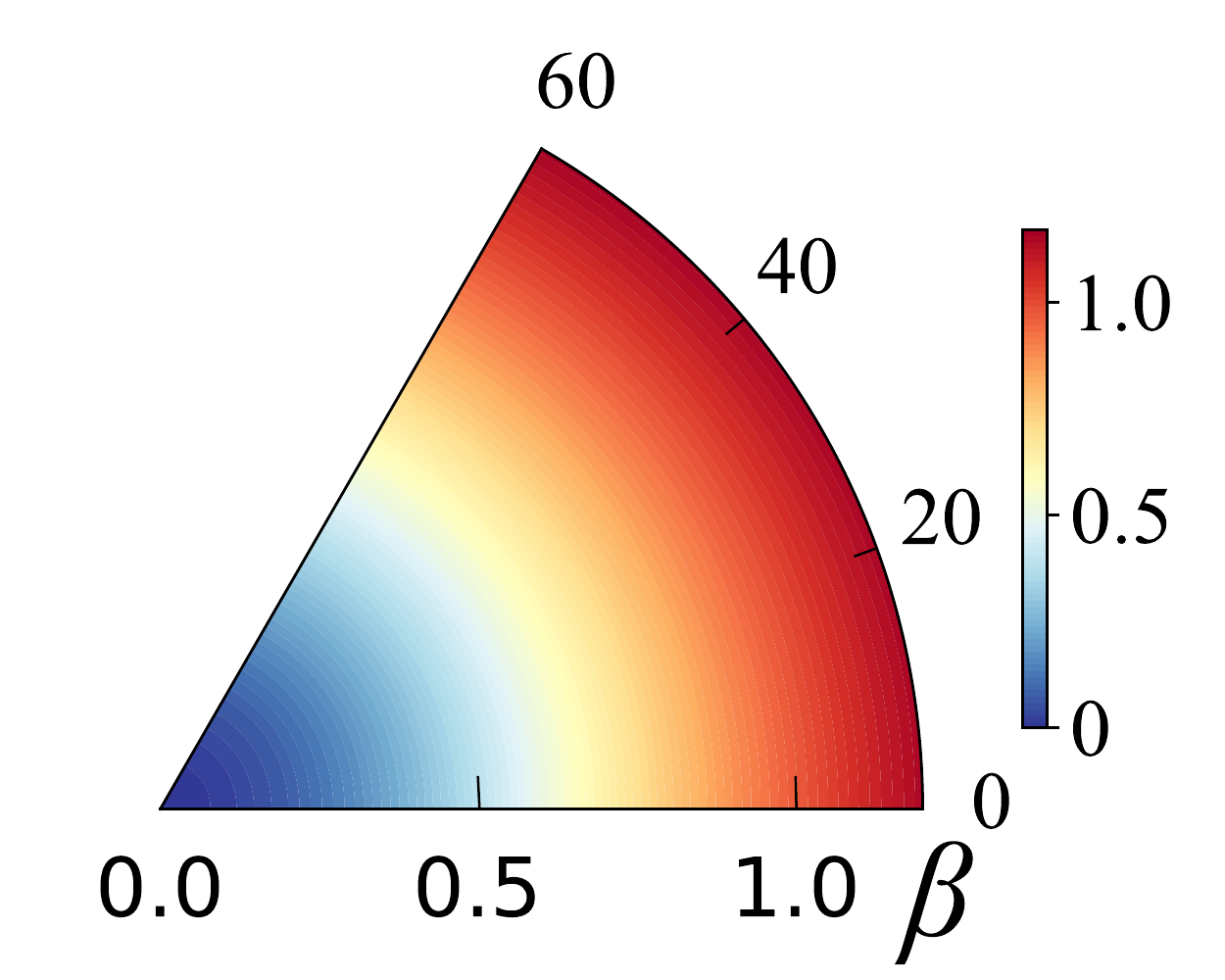}
\put (0,60) {\large $^{94}$Zr}
\end{overpic}
\begin{overpic}[width=0.19\linewidth]{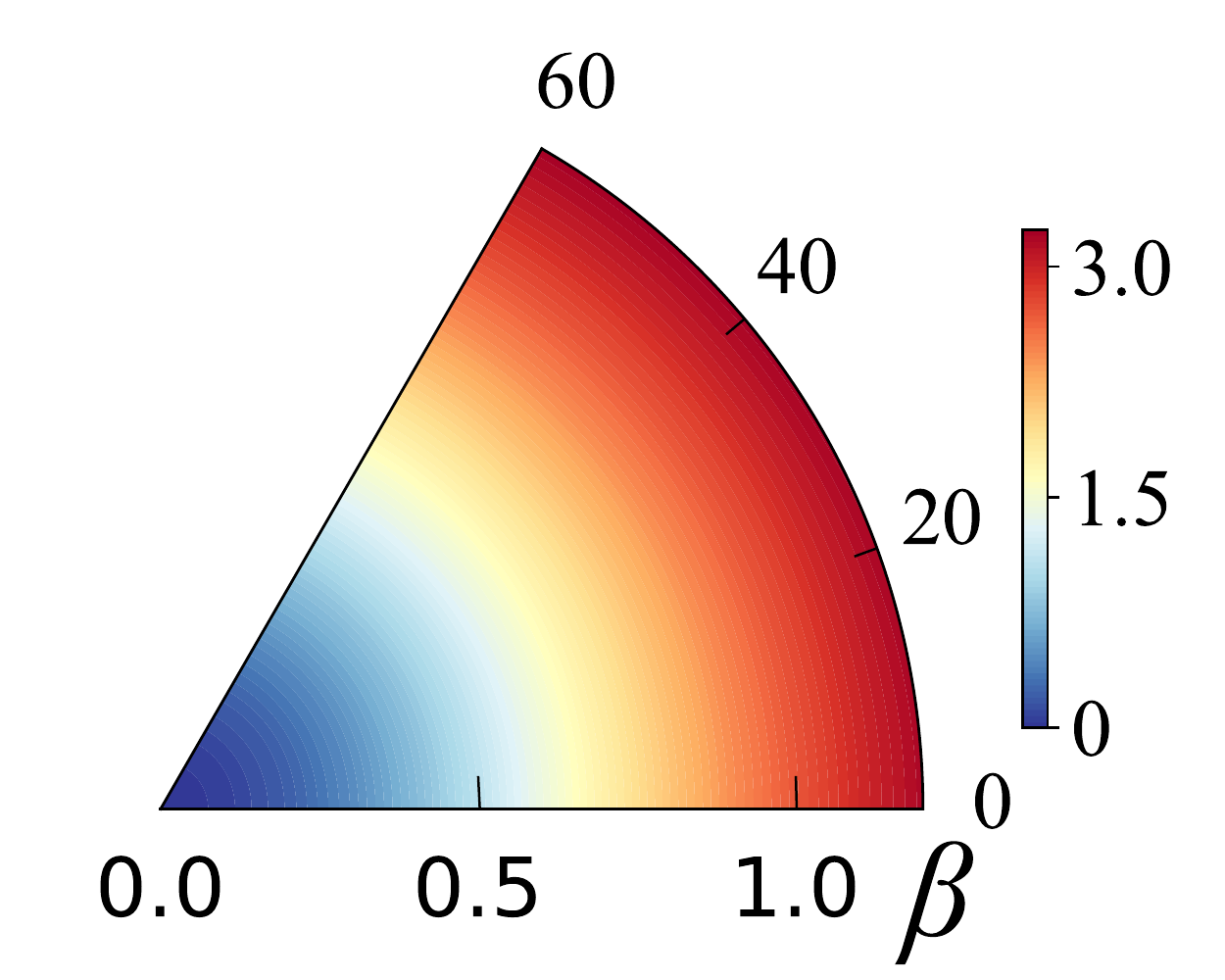}
\put (0,60) {\large $^{96}$Zr}
\end{overpic}
\begin{overpic}[width=0.19\linewidth]{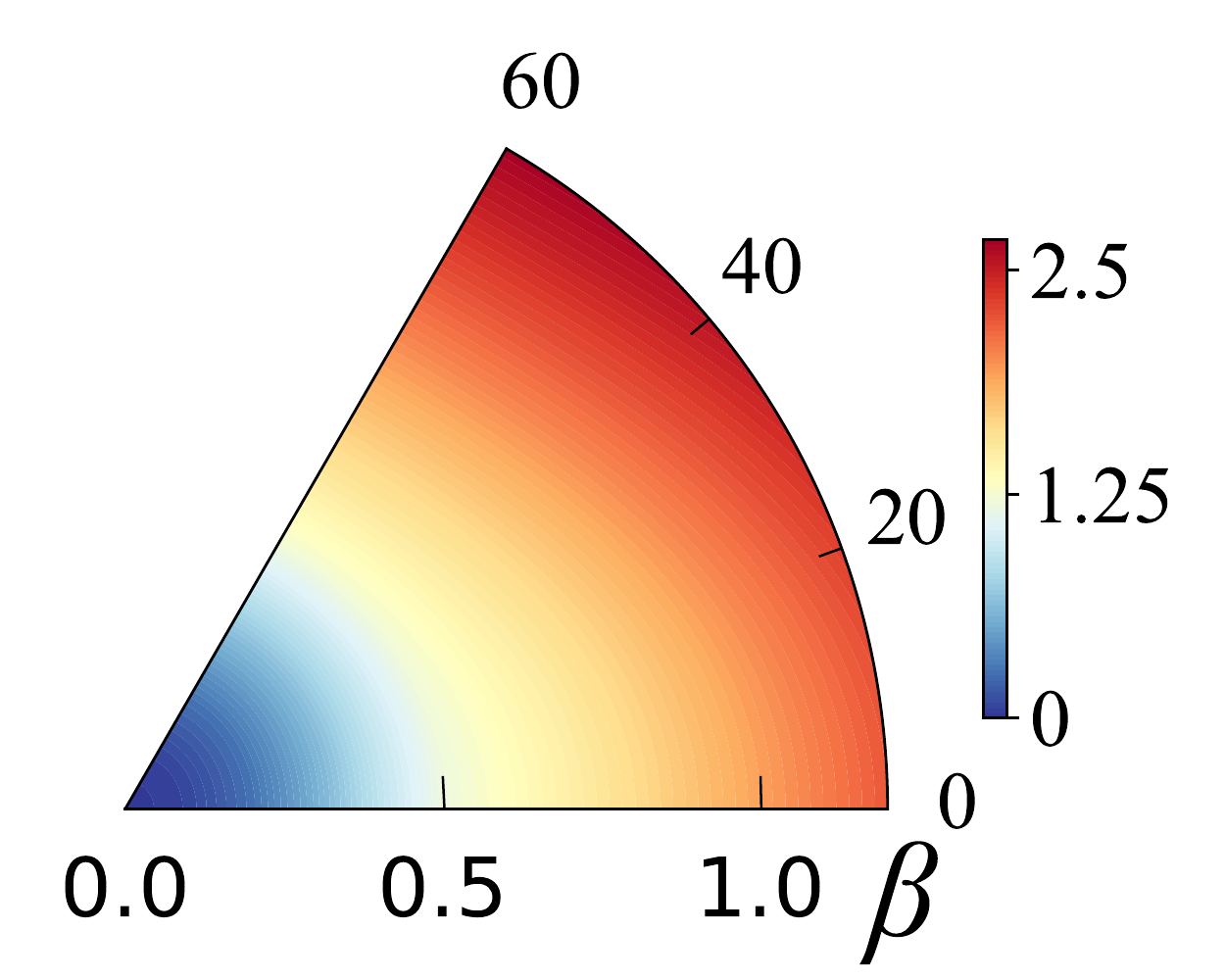}
\put (0,60) {\large $^{98}$Zr}
\end{overpic}
\begin{overpic}[width=0.19\linewidth]{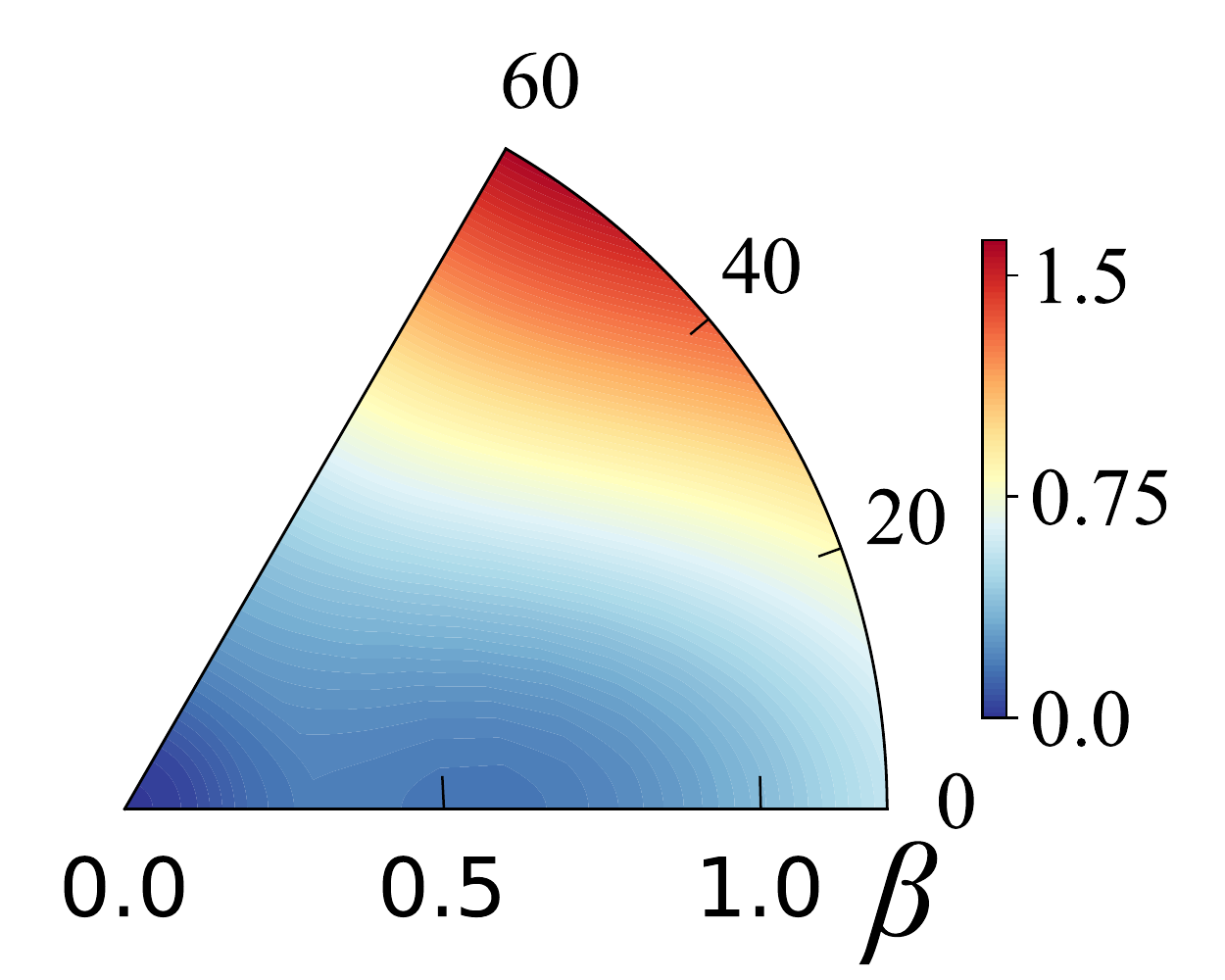}
\put (0,60) {\large $^{100}$Zr}
\end{overpic} \\ 
\begin{overpic}[width=0.19\linewidth]{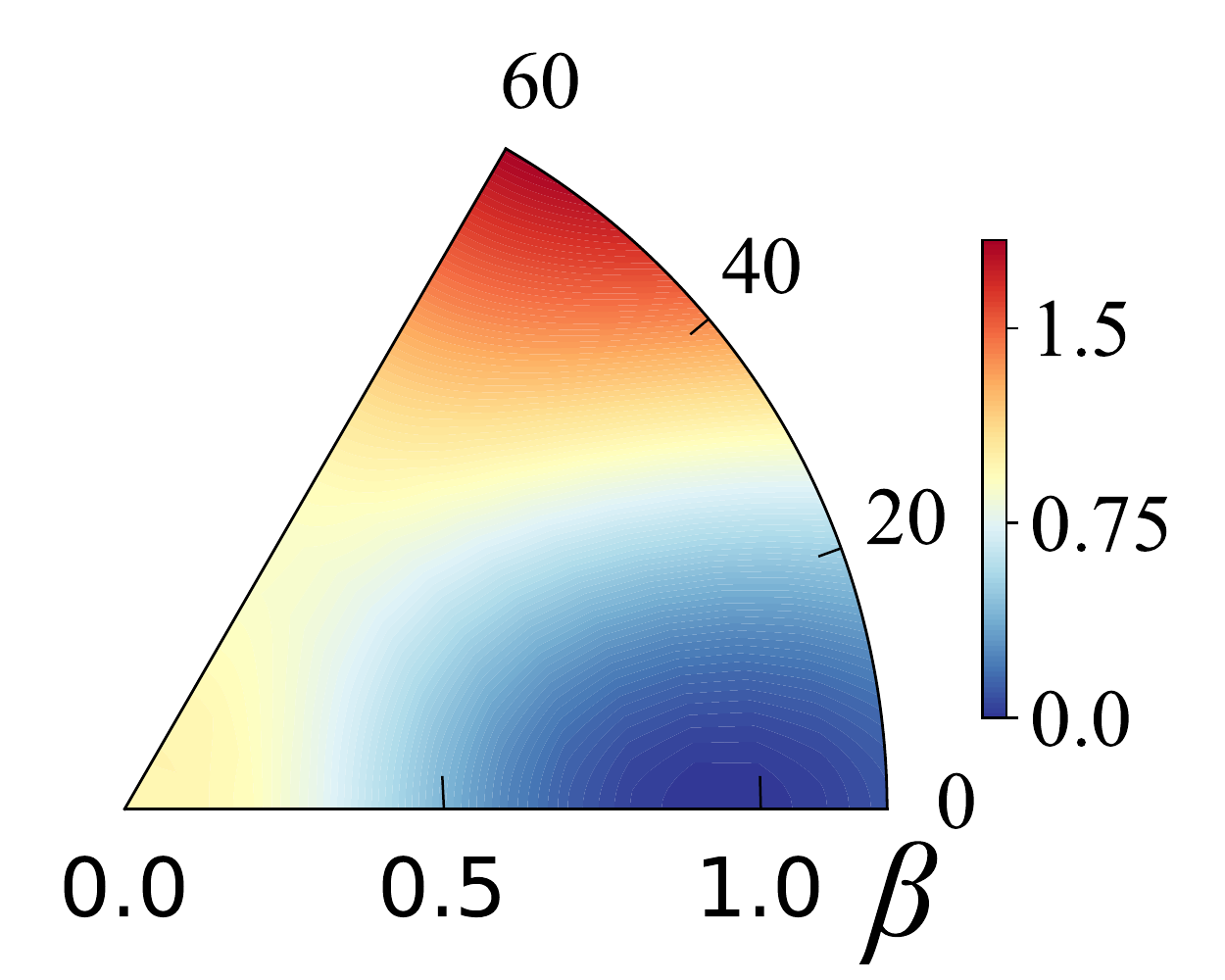}
\put (0,60) {\large $^{102}$Zr}
\end{overpic}
\begin{overpic}[width=0.19\linewidth]{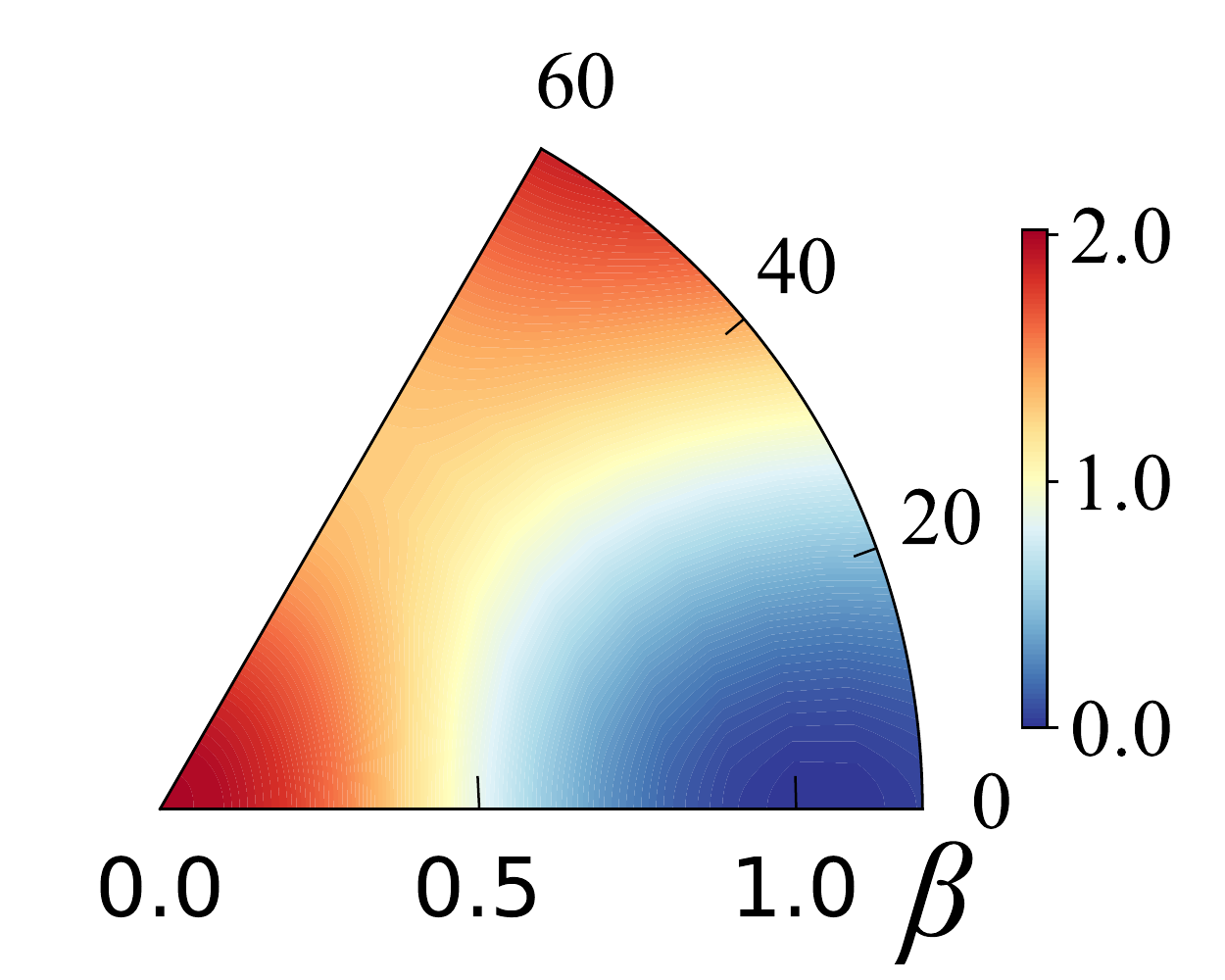}
\put (0,60) {\large $^{104}$Zr}
\end{overpic}
\begin{overpic}[width=0.19\linewidth]{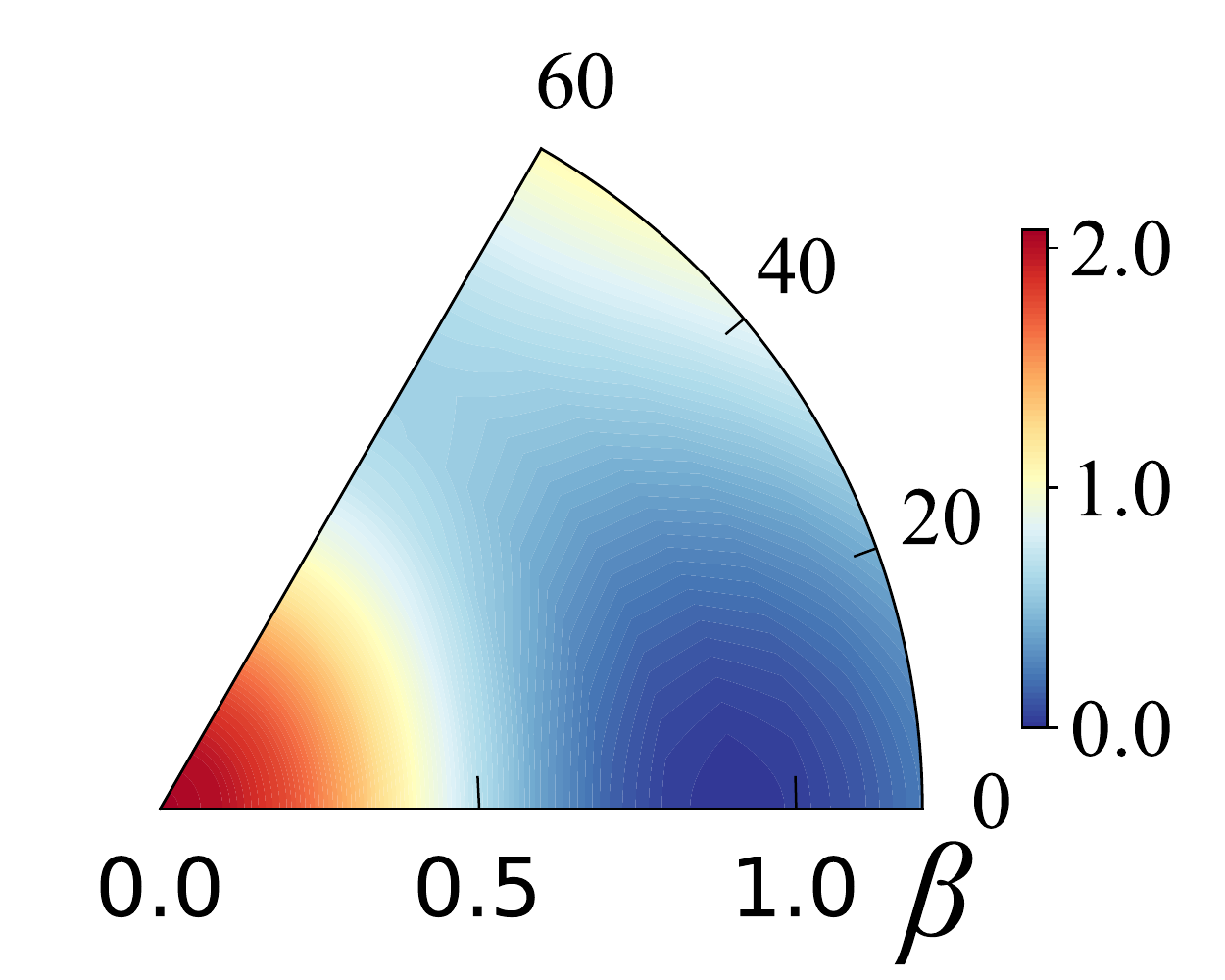}
\put (0,60) {\large $^{106}$Zr}
\end{overpic}
\begin{overpic}[width=0.19\linewidth]{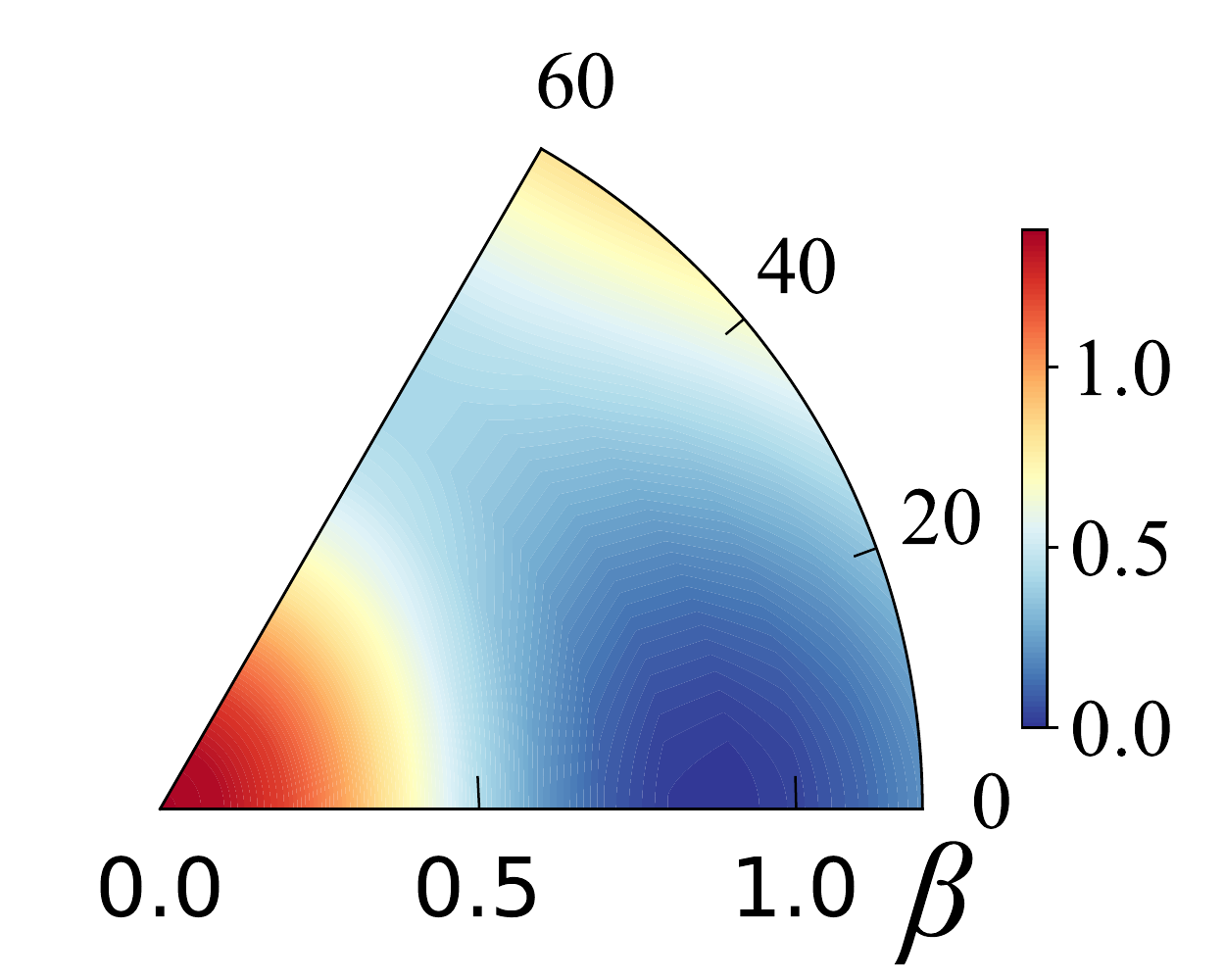}
\put (0,60) {\large $^{108}$Zr}
\end{overpic}
\begin{overpic}[width=0.19\linewidth]{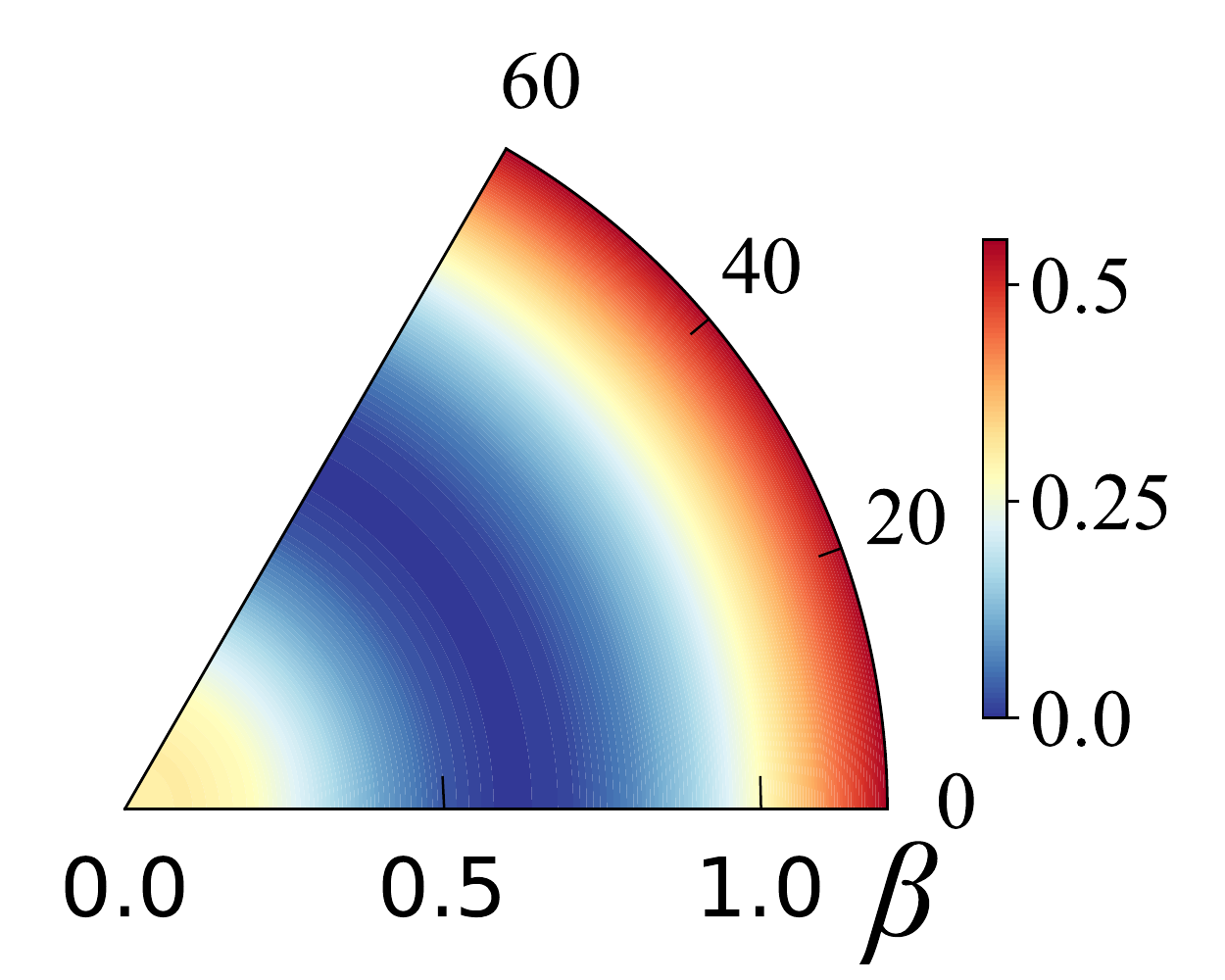}
\put (0,60) {\large $^{110}$Zr}
\end{overpic}
\caption{\label{Eminus}
\small
Contour plots in the $(\beta ,\gamma )$ plane of the lowest
eigen-potential surface, $E_{-}(\beta ,\gamma )$, for the $^{92-110}$Zr
isotopes, obtained from diagonalizing the
matrix~(\ref{eq:potential-mat}) with entries given in Eq.~(14).}
\end{figure*}
\begin{figure*}[t]
\begin{overpic}[width=1\linewidth]{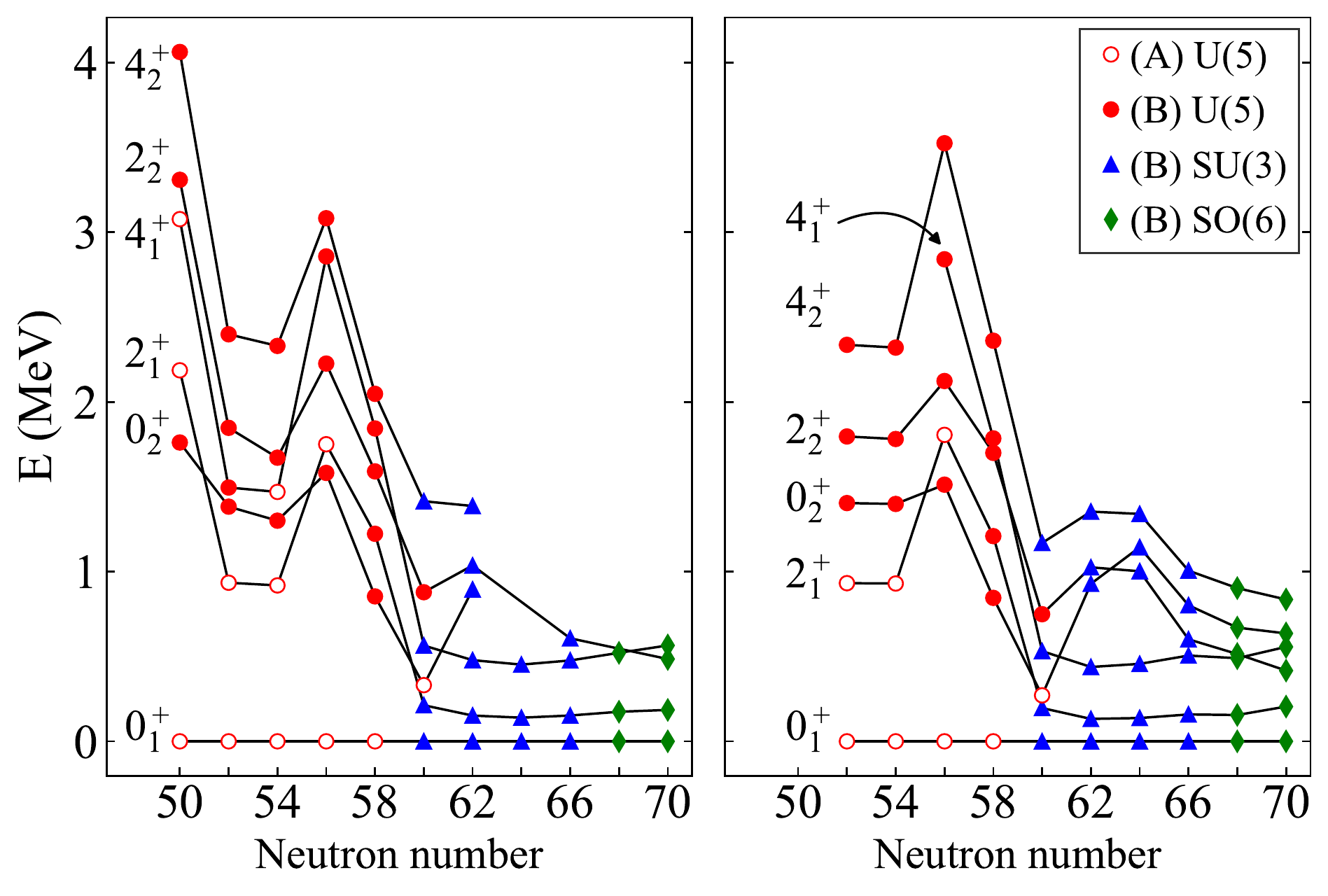}
\put (31,61) {\Large (a) {\bf Exp}}
\put (56,61) {\Large (b) {\bf Calc}}
\end{overpic}
\caption{Comparison between (a)~experimental~\cite{Paul2017,ensdf} and
(b)~calculated energy levels
$0_{1}^{+},2_{1}^{+},4_{1}^{+},0_{2}^{+},2_{2}^{+},4_{2}^{+}$.
Empty (filled) symbols indicate a state dominated by the
normal $A$-configuration (intruder $B$-configuration),
with assignments based on the decomposition of Eq.~(\ref{eq:wf}).
The shape of the symbol [$\circ,\,\triangle,\,\Diamond$], 
indicates the closest DS [U(5), SU(3), SO(6)] relevant to the
level considered. Note that the calculated values
start at neutron number 52, while the experimental values include the
closed shell at 50.
\label{fig:levels}}
\end{figure*}

The assignment of a given state to the normal $A$-configuration
or to the intruder $B$-configuration, can be inferred from
the probabilities $a^2$ or $b^2$ of the decomposition, Eq.~(\ref{eq:wf}).
The closest dynamical symmetry to the state
considered, is determined by expanding its
wave function in the U(5), SU(3) and SO(6) bases.
Fig.~\ref{fig:decomp} shows the  percentage of the wave function
within the intruder configuration for the ground
($0^{+}_1$) and excited ($0^{+}_2$) states.
The rapid change in structure of the $0^{+}_1$ state from the
normal $A$-configuration in $^{98}$Zr to the intruder
$B$-configuration in $^{100}$Zr is clearly evident.
The $0^{+}_2$ state shows a similar behavior but with the
roles of the two configurations exchanged. In $^{102}$Zr both states
belong to the intruder $B$-configuration.

One of the main advantages of the algebraic method employed,
is that one can do both a quantum and a classical analysis. 
In Fig.~\ref{Eminus}, we show the calculated 
lowest eigen-potential $E_{-}(\beta ,\gamma )$, which is the lowest
eigenvalue of the two-by-two matrix (\ref{eq:potential-mat}),
with elements given in Eq.~(14).
These classical potentials confirm the quantum results,
as they show a transition from spherical ($^{92-98}$Zr),
to a flat-bottomed potential at $^{100}$Zr,
to prolate axially-deformed ($^{102-104}$Zr), and finally to 
$\gamma$-unstable ($^{106-110}$Zr).

\section{Evolution of energy levels}

An important clue for understanding the change in structure of the Zr
isotopes, is obtained by examining the evolution of their spectra along
the chain.
In Fig.~\ref{fig:levels}, we show a comparison between experimental and 
calculated levels, along with assignments to configurations based on
Eq.~(\ref{eq:wf}), and to the closet dynamical symmetry for each level.
One can see here a rather complex structure.
In the region between neutron number 50 and 56, there appear to be two
configurations, one spherical (seniority-like), ($A$),
and one weakly deformed, ($B$), as evidenced
by the ratio $R_{4/2}$ in each configuration
which is at 52-56, $R^{(A)}_{4/2}\cong 1.6 $
and $R^{(B)}_{4/2} \cong 2.3$.
From neutron number 58, there is a pronounced drop in energy for the
states of configuration~($B$), and at 60, the two configurations
exchange their roles. This is evident in Fig.~\ref{fig:decomp} from the change in the
decomposition of the ground state $0^{+}_1$ from
configuration $A$ $(a^2 \!=\! 98.2\%)$ in $^{98}$Zr,
to configuration $B$ $(b^2 \!=\! 87.2\%)$ in $^{100}$Zr.  
The $0^{+}_2$ state displays the opposite trend, changing from
configuration $B$ in $^{98}$Zr ($b^2\!=\!98.2\%$)
to configuration $A$ ($ a^2\!=\!80.2\%$) in $^{100}$Zr. 
At this stage,
the intruder configuration~($B$) appears to be at the critical point
of a U(5)-SU(3) QPT, as evidenced in Figs.~\ref{fig:spectrum}(c)
and ~\ref{fig:spectrum}(d), by the low value of the excitation
energy of the $0^{+}_3$ state in $^{100}$Zr,
which is the first excited $0^+ $ state of the $B$-configuration
($ b^2\!=\!92.9\%$). The spectrum of states in this configuration
resembles that of the X(5) critical-point symmetry~\cite{Iac2001}.
The same situation is seen in the $_{62}$Sm
and $_{64}$Gd isotopes at neutron number 90~\cite{IBM,Scholten1978}.
In  $^{102}$Zr, that state becomes the first excited $0^{+}_2$ state
and serves as the band-head of a $\beta$-band.
Interestingly, the change in configurations appears sooner in the
$2^+_1$ level, which changes to configuration~$B$ ($b^2 \!=\! 97.1\%$)
already in $^{98}$Zr, as pointed out in~\cite{Witt2018}.
In general, beyond neutron number 60, the intruder configuration~($B$)
becomes progressively strongly deformed, as evidenced by the small
value of the excitation energy of the state $2_{1}^{+}$,
$E_{2^+_1}\!=\!151.78$ keV and by the ratio $R^{(B)}_{4/2}\!=\!3.15$
in $^{102}$Zr, and $E_{2_{1}^{+}}\!=\!139.3$~keV, $R^{(B)}_{4/2}\!=\!3.24$
in $^{104}$Zr. At still larger neutron number 66,
the ground state band becomes $\gamma $-unstable (or triaxial) as
evidenced by the close energy of the states $2_{2}^{+}$ and $4_{1}^{+}$, 
$E_{2_{2}^{+}}\!=\!607.0$~keV, $E_{4_{1}^{+}}\!=\!476.5$ keV, in $^{106}$Zr, 
and especially by the recent results 
$ E_{4^+_1}\!=\!565$~keV and $ E_{2^+_2}\!=\!485$ keV 
in $^{110} $Zr~\cite{Paul2017}, a signature of the SO(6) symmetry. 
In this region, the ground state 
configuration undergoes a~crossover from SU(3) to SO(6).

\section{Evolution of order parameters and related observables}

The above spectral analysis suggests a remarkable interplay of
configurations-interchange and shape-evolution in the Zr isotopes,
manifesting simultaneously two types of quantum phase transitions (QPTs).
The first type of QPT involves an abrupt crossing of the normal and
intruder configurations. A second type of QPT involves a gradual
shape change of the intruder configuration
which undergoes a first-order U(5) to SU(3) transition
and an SU(3) to SO(6) crossover.
In order to understand the nature of these phase transitions, 
one needs to study the behavior of the order parameters. 
In the present study, the latter involve 
the expectation value of $\hat{n}_d$ in the ground state wave function,
$\ket{\Psi; L\!=\!0^{+}_1}$ and in its
$\Psi_A$ and $\Psi_B$ components~(\ref{eq:wf}), denoted by 
$\braket{\hat{n}_d}_{0^{+}_1}$, $\braket{\hat{n}_d}_A$,
$\braket{\hat{n}_d}_B$, respectively.
As can be inferred from Eq.~(\ref{order-p}),
$\braket{\hat{n}_d}_A$ and $\braket{\hat{n}_d}_B$ portray the
shape-evolution in configuration~($A$) and ($B$), respectively.
$\braket{\hat{n}_d}_{0^{+}_1}$ involves a sum of these quantities
weighted by the probabilities of the $\Psi_A$ and and $\Psi_B$ components,
\ba
\braket{\hat{n}_d}_{0^{+}_1} =
a^2\braket{\hat{n}_d}_A + b^2\braket{\hat{n}_d}_B ~,
\label{nd-tot}
\ea
hence contains information on the normal-intruder mixing
in $\ket{\Psi; L\!=\!0^{+}_1}$.
\begin{figure*}[t]
\centering
\includegraphics[width=0.87\linewidth]{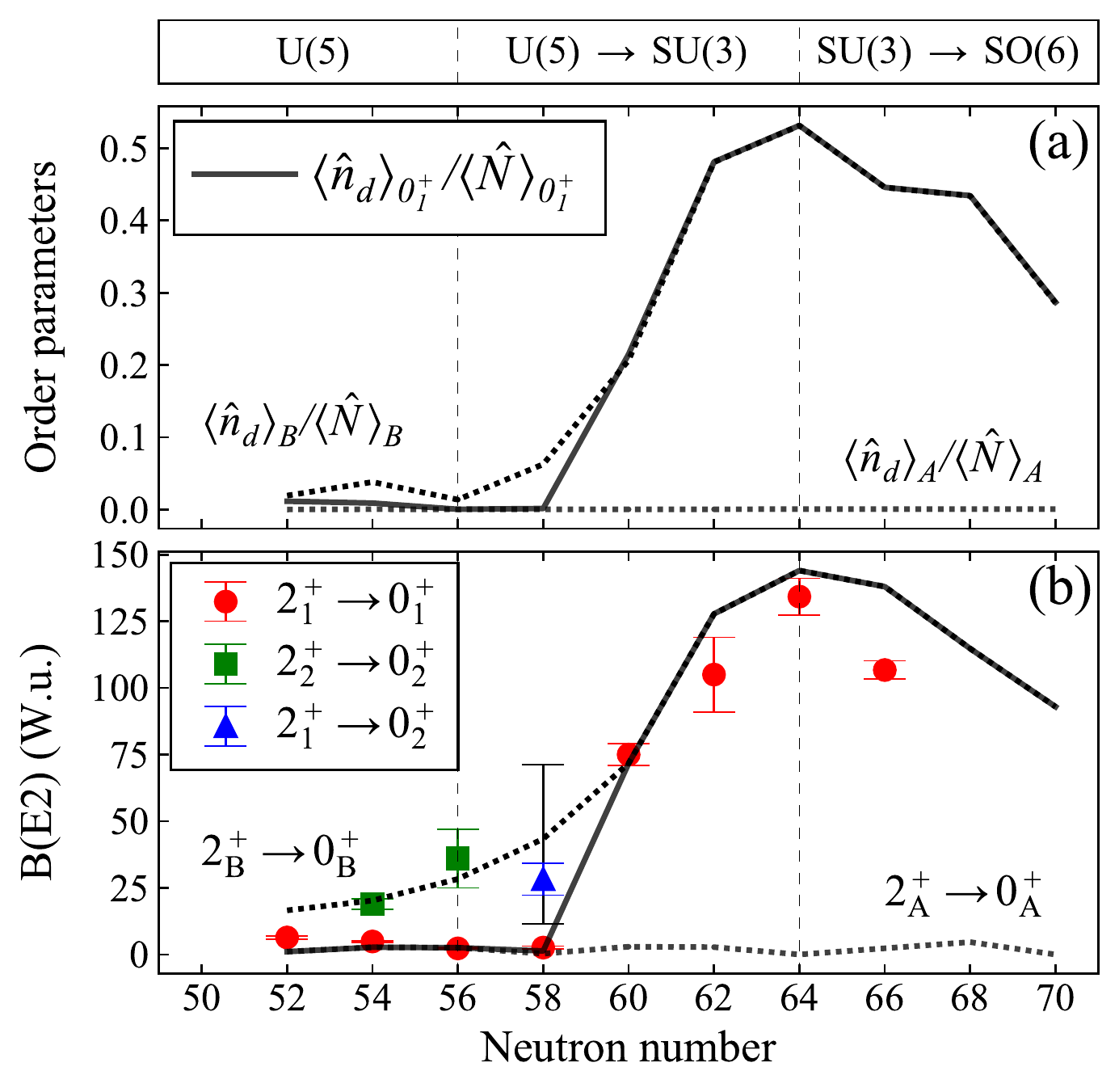}
\caption{Evolution of order parameters and of observables along the
Zr chain. Symbols (solid lines) denote
experimental data (calculated results).
(a)~The order parameters are the calculated
expectation values of $\hat{n}_d$ in the
total ground state wave function $\ket{\Psi; L=0^{+}_1}$,
Eq.~(\ref{eq:wf}) and in its ($A$) and ($B$) components
(dotted lines), normalized by the respective boson numbers.
(b)~$B(E2)$ values in Weisskopf units (W.u.).
Data taken from~\cite{Chakraborty2013, Browne2015, Kremer2016,
Ansari2017,Witt2018,Singh2018,ensdf}.
Dotted lines denote calculated $E2$ transitions within a configuration.
\label{fig:nd-be2}}
\end{figure*}
\begin{figure*}[t]
\centering
\includegraphics[width=0.86\linewidth]{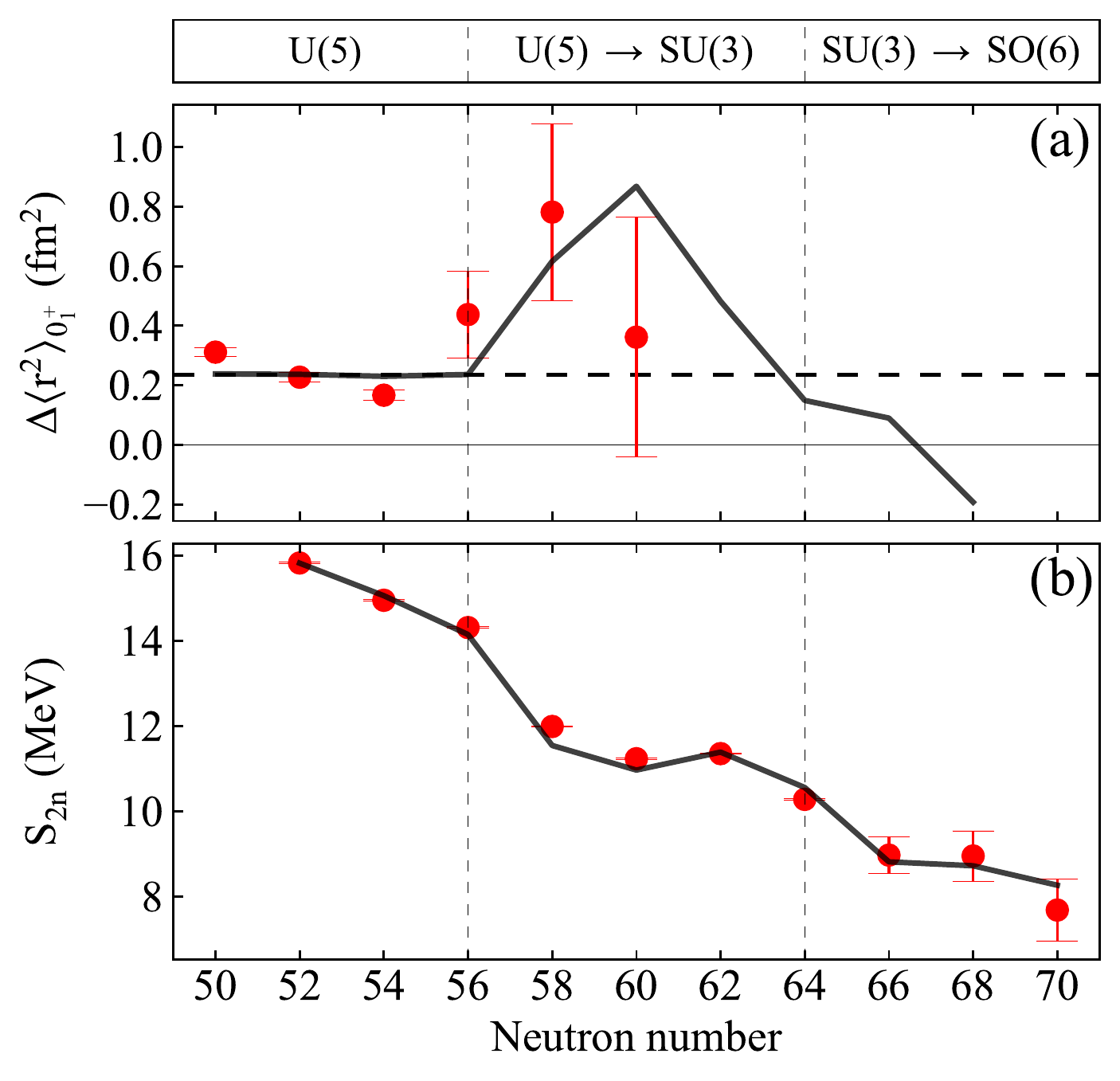}
\caption{Evolution of observables along the Zr chain.
Symbols (solid lines) denote experimental data (calculated results).
(a)~Isotope shift, $\Delta\braket{\hat{r}^{2}}_{0^{+}_1}$ in fm$^{2}$. Data
taken from~\cite{Angeli2013}. The horizontal dashed line at $0.235$
fm$^{2}$ represents the smooth behavior in
$\Delta \braket{\hat{r}^{2}}_{0^{+}_1}$ due to the $A^{1/3}$ increase of
the nuclear radius. (b)~Two-neutron separation energies,
$\rm{S}_{2n}$, in MeV. Data taken from AME2016~\cite{Wang2017}.
\label{fig:iso-S2n}}
\end{figure*}

Fig.~\ref{fig:nd-be2}(a) shows the evolution along the Zr chain 
of these order parameters 
($\braket{\hat{n}_d}_{A},\,\braket{\hat{n}_d}_{B}$ in dotted 
and $\braket{\hat{n}_d}_{0^{+}_1}$ in solid lines),
normalized by the respective boson numbers,
$\braket{\hat{N}}_A\!=\!N$, 
$\braket{\hat{N}}_B\!=\!N\!+\!2$,
$\braket{\hat{N}}_{0^{+}_1}\!=\!a^2N\!+\!b^2(N\!+\!2)$. 
Configuration~($A$) is seen to be spherical for all
neutron numbers considered. 
In contrast, configuration~($B$) is weakly-deformed
for neutron number 52-58. One can see here clearly a jump 
between neutron number 58 and 60 from
configuration~($A$) to configuration~($B$), indicating a 
first-order configuration-changing phase transition,
a further increase at neutron numbers 60-64 indicating a U(5)-SU(3)
shape-phase transition within configuration ($B$),
and, finally, there is a decrease at neutron number 66, due in part to
the crossover from SU(3) to SO(6) and in part to the shift from boson 
particles to boson holes after the middle of the major shell 50-82. 
$\braket{\hat{n}_d}_{0^{+}_1}$ is close to $\braket{\hat{n}_d}_A$
for neutron number 52-58 and coincides with $\braket{\hat{n}_d}_B$ 
at 60 and above, consistent with a high degree of purity with respect 
to configuration-mixing.

The above conclusions are stressed by an analysis of other observables, 
in particular, the $B(E2)$ values.
As shown in Fig.~\ref{fig:nd-be2}(b), the calculated $B(E2)$'s 
agree with the empirical values and follow the same trends as the 
respective order parameters. The dotted lines denote calculated
$E2$ transitions between states within the same configuration.
The $2^{+}_A\to 0^{+}_A$ transition rates
coincide with the empirical $2^{+}_1\to 0^{+}_1$
rates for neutron number 52-56. The calculated
$2^{+}_B\to 0^{+}_B$ transition rates
coincide with the empirical $2^{+}_2\to 0^{+}_2$ rates for
neutron number 52-56, with the empirical $2^{+}_1\to 0^{+}_2$ rates
at neutron number 58, and with the empirical $2^{+}_1\to 0^{+}_1$ rates
at neutron number 60-64. The large jump in $B(E2;2^+_1\rightarrow0^+_1)$
between neutron number 58 and 60 reflects the passing through a critical
point, common to a QPT involving a crossing of two configurations and
a spherical to deformed U(5)-SU(3) type of QPT within the
$B$-configuration. The further increase in $B(E2;2^+_1\rightarrow0^+_1)$
for neutron numbers 60-64 is as expected for a U(5)-SU(3) QPT
within configuration ($B$) (see Fig.~2.20 in~\cite{IBM}) and,
as in Fig.~\ref{fig:nd-be2}(a), reflects an increase in the
deformation in a spherical to deformed shape-phase transition.
The subsequent decrease from the peak at neutron number 64 towards 70,
is in accord with an SU(3) to SO(6) crossover (see Fig.~2.22
in~\cite{IBM}).

In general, the results of the current phenomenological study
resemble those obtained in the microscopic approach of the
MCSM~\cite{Togashi2016}
(which focuses on spectra and $E2$ rates), 
however, there are some noticeable differences. Specifically,  
the replacement $\gamma$-unstable $\rightarrow$ triaxial 
and the inclusion of more than two configurations in the MCSM. 
The spherical state in $^{100}$Zr is identified in the MCSM as $0^{+}_4$,
in contrast to $0^{+}_2$ in the current calculation and the data.
Both calculations show a large jump in 
$B(E2;2^+_1\rightarrow0^+_1)$, between $ ^{98} $Zr and $ ^{100} $Zr, 
typical of a first-order QPT. This is in contrast with mean-field 
based calculations~\cite{Delaroche2010, Nomura2016, Mei2012},
which due to their character smooth out the phase transitional
behavior, and show no such jump at the critical point of the QPT
(see Fig.~2 of~\cite{Singh2018}).
The observed peak in $B(E2;2^+_1\rightarrow0^+_1)$ for $^{104}$Zr,
is reproduced by the current calculation but not by the MCSM.

Further evidence for the indicated structural changes occurring in
the Zr chain, can be obtained from an analysis of the isotope shift 
$\Delta\braket{\hat r^2}_{0^+_1}
=\braket{\hat{r}^{2}}_{0^{+}_1;A+2}-\braket{\hat{r}^2}_{0^{+}_1;A}$, where 
$\braket{\hat r^2}_{0^+_1} $ is the expectation value of $ \hat r^2 $
in the ground state $ 0^+_1 $. In the IBM-CM the latter is given by 
\ba
\label{eq:charge-rad}
\braket{\hat r^2} = r^2_c + \alpha N_v + \eta [\braket{\hat n_d^{(N)}} 
  + \braket{\hat n_d^{(N+2)}}] ~,
\ea
where $r^2_c$ is the square radius of the closed shell, 
$N_v$ is half the number of valence particles, 
and $\eta$ is a coefficient that takes into account the effect 
of deformation\cite{IBM,Zerguine2008,Zerguine2012}.
The isotope shift depends on two parameters,
$\alpha\!=\!0.235,\,\eta \!=\! 0.264$ fm$^2$, whose values are fixed
by the procedure of Ref.~\cite{Zerguine2008,Zerguine2012}.
$\Delta\braket{\hat r^2}_{0^+_1}$ should increase at the transition
point and decrease and, as seen in Fig.~{\ref{fig:iso-S2n}(a),
it does so, although the error bars are large and no data are available
beyond neutron number 60. (In the large $N$ limit, this quantity, 
proportional to the derivative of the order parameter 
$\braket{\hat{n}_d}_{0^{+}_1}$, diverges at the critical point).

Similarly, the two-neutron separation energies $S_{2n} $ 
can be written as~\cite{IBM},
\ba
S_{2n} = -\tilde{A} -\tilde{B} N_v \pm S^{\rm def}_{2n} - \Delta_n ~,
\ea
where $S^{\rm def}_{2n}$ is the contribution of the deformation,
obtained by the expectation value of the Hamiltonian in the
ground state~$ 0^+_1$.
The $ + $ sign applies to particles and the $ - $ sign to holes,
and $\Delta_n $ takes into account the neutron subshell closure at 56, 
$\Delta_n = 0 $ for 50-56 and $ \Delta_n = 2 $ MeV for 58-70.
The value of $ \Delta_n $ is taken from Table XII of~\cite{Barea2009} 
and $ \tilde{A}\!=\!-16.5,\,\tilde{B}\!=\!0.758$ MeV are determined
by a fit to binding energies of $^{92,94,96}$Zr.
The calculated $ S_{2n}$, shown in Fig.~\ref{fig:iso-S2n}(b),
displays a complex behavior. Between neutron number 52 and 56 
it is a straight line, as the ground state is spherical (seniority-like)
configuration~($A$). After 56, it first goes down due to the subshell
closure at~56, then it flattens as expected from a first-order QPT 
(see, for example the same situation in the $_{62}$Sm
isotopes~\cite{Scholten1978}). After 62, 
it goes down again due to the increasing of deformation and finally it
flattens as expected from a crossover from SU(3) to SO(6).

\section{Conclusions}

We have presented here a quantum analysis of spectra
and other observables (including $E2$ rates, isotope shifts,
separation energies) and a classical analysis of shapes, 
for the entire chain of $_{40}$Zr isotopes, from neutron number 52 to 70.
The calculations were performed within the IBM-CM, which
provides a simple tractable shell-model-inspired algebraic framework,
where global trends of structure and symmetries can be clearly identified
and diversity of observables calculated. The evolution of structure and
QPT attributes, along the Zr chain, are studied by varying the control
parameters in the IBM-CM Hamiltonian followed by a detailed comparison
with the available experimental data on yrast and non-yrast states.

The results of the comprehensive analysis suggest a complex phase
structure in these isotopes, involving two configurations.
The normal $A$ configuration remains spherical in all isotopes considered.
The intruder $B$-configuration undergoes first a spherical to
axially-deformed U(5)-SU(3) QPT, with a critical-point near $^{100}$Zr,
and then an axially-deformed to $\gamma$-unstable SU(3)-SO(6) crossover.
In parallel to the gradual shape-evolution within configuration $B$,
the two configurations cross near neutron number 60, and the ground state
changes from configuration ($A$) to configuration ($B$).
Interestingly, the critical-point of the U(5)-SU(3) shape-changing QPT
coincides with the critical-point of the configuration-changing QPT.
The two configurations are weakly mixed and retain their purity
before and after the crossing.

Further details of our results, including the calculation of spectra and 
transition rates in all the $^{92-110}$Zr isotopes and of other quantities 
not reported here, will be given in a forthcoming publication
based on~\cite{gavrielov-thes}. 
Our method of calculation could also be applied to the $_{38}$Sr
isotopes, which show similar features~\cite{Mach1989}.
The present work provides the first evidence for
multiple quantum phase transitions in nuclear physics
and may stimulate research for this type of phase transitions in other
fields of physics.

\ack
This work was supported in part by U.S. DOE under Grant No.
DE-FG02-91ER-40608 and by the US-Israel Binational Science Foundation 
Grant No. 2016032. We thank R.~F. Casten and J.~E.~Garc\'\i a-Ramos for
insightful discussions and for bringing to our attention
a recently reported IBM-CM calculation for the Zr isotopes \cite{Ramos2019},
which conforms with the results of the present analysis.\\

\providecommand{\newblock}{}


\begin{thebibliography}{10}
\expandafter\ifx\csname url\endcsname\relax
\def\url#1{{\tt #1}}\fi
\expandafter\ifx\csname urlprefix\endcsname\relax\def\urlprefix{URL }\fi
\providecommand{\eprint}[2][]{\url{#2}}

\bibitem{Cheifetz1970}
Cheifetz E R, Jared C, Thompson S G and J. B. Wilhelmy J B 1970,
{\rm Phys. Rev. Lett.} {\bf 25} 38

\bibitem{Federman1977}
Federman P and Pittel S 1977
{\em Phys. Lett.\/} B {\bf 69} 385

\bibitem{Federman1979}
Federman P and Pittel S 1979
{\em Phys. Rev.\/} C {\bf 20} 820

\bibitem{HeydeCas1985}
Heyde K, {Van Isacker} P, Casten R~ and Wood J~L 1985
  {\em Phys. Lett.\/} B {\bf 155} 303

\bibitem{Heyde2011}
Heyde K and Wood J~L 2011 {\em Rev. Mod. Phys.\/} {\bf 83} 1467

\bibitem{Delaroche2010}
  Delaroche J~P, Girod M, Libert J, Goutte H, Hilaire S,
  P{\'{e}}ru S, Pillet N and Bertsch G~F 2010
  {\em Phys. Rev.\/} C {\bf 81} 014303

\bibitem{Nomura2016}
  Nomura K, Rodr{\'{i}}guez-Guzm{\'{a}}n R and Robledo L~M 2016
  {\em Phys. Rev.\/} C {\bf 94} 044314

\bibitem{Mei2012}
  Mei H, Xiang J, Yao J~M, Li Z~P and Meng J 2012
  {\em Phys. Rev.\/} C {\bf 85} 034321

\bibitem{Langanke2009}
  Sieja K, Nowacki F, Langanke K, Mart\'inez-Pinedo G 2009
  {\em Phys. Rev.\/} C {\bf 79} 064310

\bibitem{Petrovici2012}
Petrovici A 2012
{\em Phys. Rev.\/} C {\bf 85} 034337

\bibitem{Togashi2016}
  Togashi T, Tsunoda Y, Otsuka T and Shimizu N 2016
  {\em Phys. Rev. Lett.\/} {\bf 117} 172502

\bibitem{Chakraborty2013}
  Chakraborty A {\em et~al.\/} 2013
  {\em Phys. Rev. Lett.\/} {\bf 110} 022504

\bibitem{Browne2015}
  Browne F {\em et~al.\/} 2015
  {\em Phys. Lett.\/} B {\bf 750} 448

\bibitem{Kremer2016}
  Kremer C {\em et~al.\/} 2016
  {\em Phys. Rev. Lett.\/} {\bf 117} 172503

  \bibitem{Ansari2017}
  Ansari S {\em et~al.\/} 2017
  {\em Phys. Rev.\/} C {\bf 96} 054323

\bibitem{Paul2017}
  Paul N {\em et~al.\/} 2017
  {\em Phys. Rev. Lett.\/} {\bf 118} 032501

\bibitem{Witt2018}
  Witt W {\em et~al.\/} 2018
  {\em Phys. Rev.\/} C {\bf 98} 041302

\bibitem{Singh2018}
  Singh P {\em et~al.\/} 2018
  {\em Phys. Rev. Lett.\/} {\bf 121} 192501

\bibitem{Gilmore1978} 
R. Gilmore R  and D.~H. Feng D H 1978 
{\em Phys. Lett.\/} B {\bf 76} 26

\bibitem{Gilmore1979} 
R. Gilmore R 1979 
{\em J. Math. Phys.} {\bf 20}  891

\bibitem{Carr} 
See, for example, \textit{Understanding Quantum Phase Transitions}, 
L. Carr, ed., (CRC Press, Boca Raton, FL, 2011).
  
\bibitem{CejJolCas2010}
  Cejnar P, Jolie J and Casten R~F 2010
  {\em Rev. Mod. Phys.\/} {\bf 82} 2155

\bibitem{Heyde2004}
 Heyde K, Jolie J, Fossion R, {De Baerdemacker} S and Hellemans V 2004
{\em Phys. Rev.\/} C {\bf 69} 54304

 \bibitem{GavLevIac2019}
  Gavrielov N. Leviatan A, Iachello F 2019
  {\em Phys. Rev.\/} C {\bf 99} 064324

 \bibitem{IBM}
  Iachello F and Arima A 1987
  {\em {The Interacting Boson Model}\/} (Cambridge:
  Cambridge University Press)

\bibitem{IacTal1987}
Iachello F and Talmi I 1987
{\em Rev. Mod. Phys.} {\bf 59} 339  

\bibitem{GinKir1980}
  Ginocchio J~N and Kirson M~W 1980
  {\em Phys. Rev. Lett.\/} {\bf 44} 1744

\bibitem{Diep1980}
  Dieperink A~E~L, Scholten O and Iachello F 1980
  {\em Phys. Rev. Lett.\/} {\bf 44} 1747

  \bibitem{CejJol2009} 
Cejnar P and Jolie J 2009
{\em Prog. Part. Nucl. Phys.} {\bf 62} 210

\bibitem{Iac2011} 
Iachello F 2011 
{\em Rivista del Nuovo Cimento} {\bf 34} 617

\bibitem{Warner1983}
Warner D~D and Casten R~F 1983
{\em Phys. Rev. C\/} {\bf 28} 1798

\bibitem{Lipas1985}
Lipas P O, Toivonen P and Warner D S 1985
{\em Phys. Lett.\/} B {\bf 155} 295

\bibitem{Duval1981}
Duval P~D and Barrett B~R 1981 {\em Phys. Lett.\/} B {\bf 100} 223

\bibitem{Duval1982}
Duval P~D and Barrett B~R 1982 {\em Nucl. Phys.\/} A {\bf 376} 213

\bibitem{Sambataro1982}
  Sambataro M and Moln{\'{a}}r G 1982
  {\em Nucl. Phys.\/} A {\bf 376} 201

\bibitem{Duval1983}
  Duval P~D, Goutte D and Vergnes M 1983
  {\em Phys. Lett.\/} B {\bf 124} 297

\bibitem{Bijker2006}
  Padilla-Rodal E, Casta{\~{n}}os O, Bijker R and Galindo-Uribarri A 2006
  {\em Rev. Mex. Fis.} {\bf S 52} 57

\bibitem{Fossion2003}
  Fossion R, Heyde K, Thiamova G and {Van Isacker} P 2003
  {\em Phys. Rev.\/} C {\bf 67} 024306

\bibitem{Frank2006}
  Frank A, Van Isacker P and Iachello F 2006
  {\em Phys. Rev.\/} C {\bf 73} 061302

\bibitem{Ramos2011}
Garc\'\i a-Ramos J~E, Hellemans V and Heyde K 2011
  {\em Phys. Rev.\/} C {\bf 84} 14331

\bibitem{Ramos2014}
  Garc\'\i a-Ramos J~E and Heyde K 2014
  {\em Phys. Rev.\/} C {\bf 89} 14306

\bibitem{Ramos2015}
  Garc\'\i a-Ramos J~E and Heyde K 2015
  {\em Phys. Rev.\/} C {\bf 92} 034309

\bibitem{Nomurajpg2016}
  K. Nomura, T, Otsuka and P. Van Isacker 2016
{\em J. Phys.} G {\bf 43} 024008

\bibitem{LevGav2018}
Leviatan A, Gavrielov N, J.~E.~Garc\'\i a-Ramos J~E and Van Isacker P 2018
{\em Phys. Rev.\/} C {\bf 98}  031302(R)

\bibitem{Frank2004}
  Frank A, {Van Isacker} P and Vargas C~E 2004
  {\em Phys. Rev.\/} C {\bf 69} 034323

\bibitem{Hellemans2007}
Hellemans V, Van Isacker P, De Baerdemacker S and Heyde K 2007
{\em Nucl. Phys.\/} A {\bf 789} 164

  \bibitem{Heyde1987}
  Heyde K, Jolie J, Moreau J, Ryckebusch J, Waroquier M, Duppen P~V,
  Huyse M and Wood J~L 1987
  {\em Nucl. Phys.\/} A {\bf 466} 189

\bibitem{gavrielov-thes}
  Gavrielov N,
  Ph.D. Thesis, The Hebrew University, Jerusalem, Israel (unpublished)

\bibitem{ensdf}
{Evaluated Nuclear Structure Data File (ENSDF)}
\href{https://www.nndc.bnl.gov/ensdf/}{https://www.nndc.bnl.gov/ensdf/}

\bibitem{Iac2001}
  Iachello F 2001
  {\em Phys. Rev. Lett.\/} {\bf 87} 052502

\bibitem{Scholten1978}
Scholten O, Iachello F and Arima A 1978 {\em Ann. Phys.\/} {\bf 115} 325

\bibitem{Angeli2013}
Angeli I and Marinova K~P 2013
{\em At. Data Nucl. Data Tables\/} {\bf 99} 69

\bibitem{Wang2017}
  Wang M, Audi G, Kondev F~G, Huang W, Naimi S and Xu X 2017
  {\em Chinese Phys.\/} C {\bf 41} 030003

\bibitem{Zerguine2008}
  Zerguine S, {Van Isacker} P, Bouldjedri A and Heinze S 2008
  {\em Phys. Rev. Lett.\/} {\bf 101} 022502

\bibitem{Zerguine2012}
  Zerguine S, {Van Isacker} P and Bouldjedri A 2012
  {\em Phys. Rev.\/} C {\bf 85} 034331

\bibitem{Barea2009}
  Barea J and Iachello F 2009
  {\em Phys. Rev.\/} C {\bf 79} 044301

\bibitem{Mach1989}
  Mach H, Moszynnski M, Gill R, Wohn F, Winger J, {C Hill} J,
  Moln{\'{a}}r G and Sistemich K 1989
  {\em Phys. Lett.\/} B {\bf 230} 21
  
\bibitem{Ramos2019}
	J. E. Garc\'ia Ramos and K. Heyde 2019 arXiv:1909.00824 [nucl-th]
\end{thebibliography}
\end{document}